\documentclass[12pt,preprint]{aastex}

\usepackage{amsmath, amsthm, amssymb}
\usepackage{natbib}
\usepackage{amssymb}
\usepackage{graphicx}
\usepackage{multirow}
\usepackage[normalem]{ulem}
\usepackage{color}
\usepackage{arydshln}
\usepackage{diagbox}

\newcommand {\swift} {{\it Swift} }
\newcommand {\Fermi} {Fermi }
\newcommand {\f} {$f_{NC}$ }
\newcommand{\sm} {\footnotesize }

\title{Short vs Long and  Collapsars vs. non-Collapsar: a quantitative classification of GRBs.}
\author{\large Omer Bromberg$^1$, Ehud Nakar$^2$, Tsvi Piran$^1$, Re'em Sari$^1$\\
\footnotesize $^1$ Racah Institute of Physics, The Hebrew University, 91904 Jerusalem, Israel\\
\footnotesize $^2$ The Raymond and Berverly Sackler School of Physics and Astronomy,\\
\footnotesize Tel Aviv University, 69978 Tel Aviv, Israel}

\begin{abstract}
Gamma-Ray Bursts (GRBs) are traditionally divided to long 
and short 
according to their durations ($\lessgtr 2 $ sec). It was generally
believed that this reflects a different physical origin: Collapsars
(long) and non-Collapsars (short).  We have recently shown
that  the duration distribution of Collapsars is flat, namely  independent of the duration, at short durations. Using this model for
the distribution of Collapsars we determine
the duration distribution of non-Collapsars and estimate the probability
that a burst with a given duration (and hardness) is a Collapsar or
not. We find that this probability depends strongly on the spectral
window of the observing detector. While the commonly used limit of 2
sec is conservative and suitable for BATSE bursts, 40\% of  \swift's
bursts shorter than 2 sec are Collapsars and division $\lessgtr 0.8$ sec
is more suitable for \swift. We find that the duration overlap of
the two populations is very large. On the one hand there is a
non-negligible fraction of non-Collapsars longer than 10 sec, while
on the other hand  even bursts shorter than 0.5 sec in the \swift
sample have a non-negligible probability to be Collapsars.
Our results enable the construction of non-Collapsar
samples while controlling the Collapsar contamination. They also
highlight that no firm conclusions can be drawn based on a
single burst and they have numerous implications concerning previous
studies of non-Collapsar properties that were based on the current significantly
contaminated \swift samples of localized short GRBs.
Specifically: (i)  all known short bursts with $z>1$ are most likely Collapsars,
(ii) the only short burst with a clear jet break  is most likely a Collapsar, indicating
our lack of knowledge concerning non-Collapsar beaming (iii) the existence of non-Collapsars with
durations up to 10 sec impose new challenges to non-Collapsar models.
 \end{abstract}

\begin{document}

\section{Introduction}

\citet{Kouveliotou93} have shown that  gamma ray bursts (GRBs) can
be divided to two groups according to their observed duration. Long
bursts (LGRBs) with observed durations $T_{90} >2$ sec   and short
ones  (SGRBs) with  $T_{90} <2$ sec.  They have also found that
SGRBs are harder on average than LGRBs, supporting further the
possibility that the two populations arise from different physical
sources. Later on afterglow observations enabled the localizations
of GRBs and identifications of their hosts. These observations
supported further the different sources hypothesis. Hosts of LGRBs
have a large star formation rate while SGRB hosts include both star
forming and non-star forming galaxies. The position distribution of
LGRBs within their host, towards the center and within high star
forming regions \citep{Fruchter06}, differs from the position distribution of SGRBs
within their hosts, which is more diffuse and with no apparent
association with star formation
\citep{Barthelmy05,Fox05,Gehrels05,Nakar07,Berger09}.

These observations have led to the realization that  GRBs  have two
different progenitors and they are generated by at least two
different mechanisms\footnote{We have recently shown
that low luminosity GRBs are a third group, generated by a different mechanism
than regular GRBs \citep{B11b}.}.
The association of LGRBs with star forming regions and in several
cases with type Ic SNe suggest that they involve stellar collapse.
The Collapsar model \citep{Paczynski98,MacWoos99} suggests that  a central engine within the
collapsing star (powered most likely by an accretion disk onto the
newly formed compact object or by a magnetar) powers a jet that
penetrates the stellar envelope and produces the observed gamma-rays
once it is outside the star. SGRBs are typically weaker and are
observed at lower distances. They are more numerous (locally), but
being harder to detect there are less observed SGRBs than long one.
Those SGRBs with identified locations are
associated with a wide range of stellar population ages. Their properties are consistent with those expected from  binary neutron star mergers  \citep[][]{Eichler89}, although the
exact origin is still uncertain \citep[see][for a recent review]{Nakar07}. Since the origin of this group is still
uncertain we will denote them  simply as non-Collapsars.

\begin{figure}
\includegraphics[width=7in]{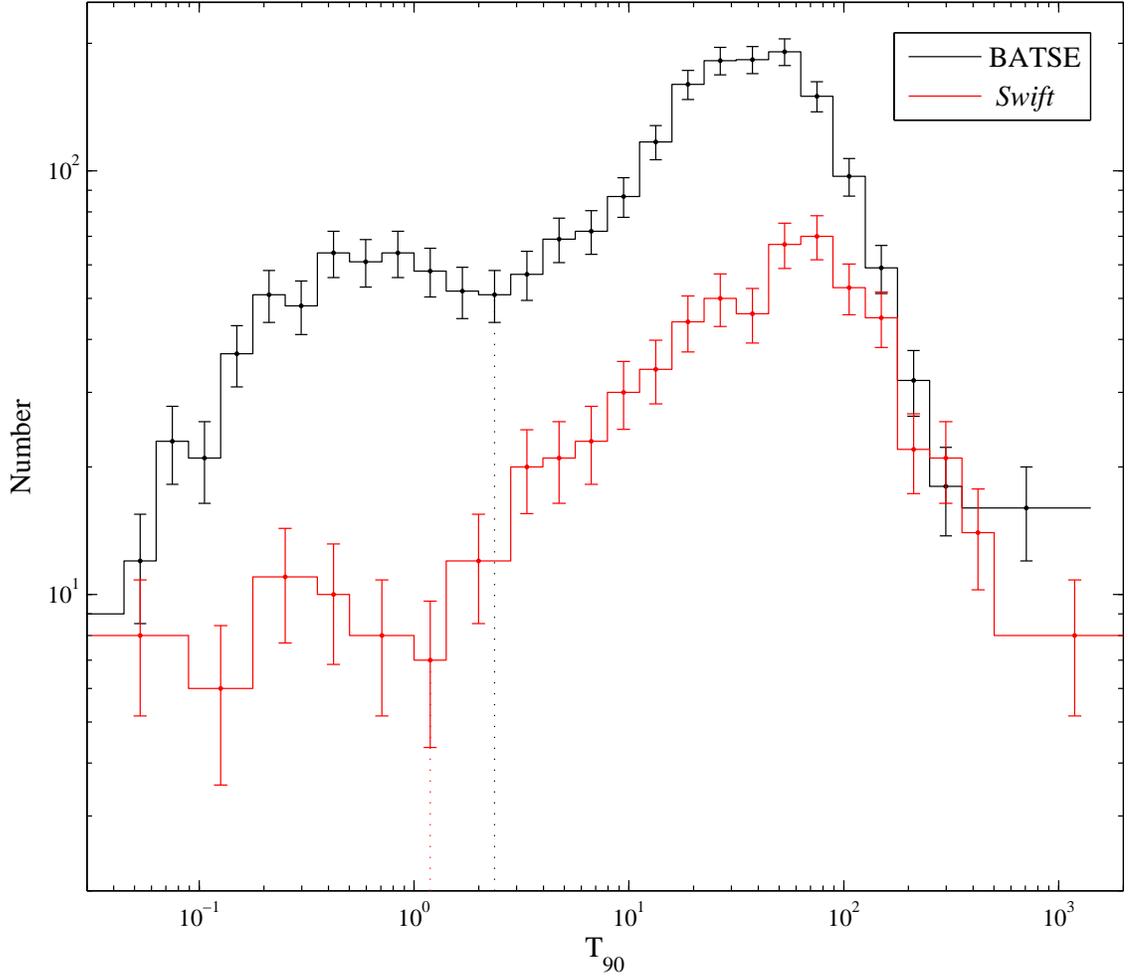}
  \caption{
  The $log(T_{90})$ double humped durations distributions, $dN/d\log (T_{90})$, of BATSE (black) and
    {\it Swift} (red), binned into equally spaced logarithmic bins.
  Bins with less than 5 events are merged with their neighbors to reduce statistical errors.
  The minima in the distributions occur at $T_{90}=2.4\pm0.4$ sec in the
  BATSE distribution and at $T_{90}=1.2\pm0.2$ sec in \swift.}
  \label{fig.dNdlogT}
\end{figure}

It is commonly implicitly assumed that there is one to one
correspondence between the observed groups of LGRBs and SGRBs and
the astrophysical groups of Collapsars and non-Collapsars. However,
a quick inspection of the duration distribution (fig.
\ref{fig.dNdlogT}) suggests that  this is not the case and there is
a significant overlap between the two groups of long and short GRBs:
there are SGRBs of Collapsar origin and vice versa.
Apart from a slight difference in the average hardness, all other
high energy emission properties of LGRBs and SGRBs are remarkably
similar. This makes it difficult to identify the origin  of any
individual burst \citep{Nakar07} and lacking a better criteria the
original division according to  $T_{90} \lessgtr 2 $ sec is widely
used. However this criteria  was established for a specific detector
(BATSE) with a specific observational window. SGRBs are typically
harder than long ones and as such they are more difficult to detect
by
softer detectors like \swift/BAT than by BATSE. Indeed, {\it
Swift}'s   short/long detection rate is 1/10 vs. BATSE's 1/3 (when
the criterion $T_{90} \lessgtr 2 $ is used). This suggests that {\it
Swift}'s division line between the two groups might be at a shorter
duration, as indeed is seen in a visual inspection of {\it Swift}'s
duration distribution (see fig. \ref{fig.dNdlogT}).

\citet[][]{Zhang09} suggested to classify individual GRBs using
various subsets of  properties, e.g. spectral lag, peak energy, etc.
These subsets of properties  are selected phenomenologically, and
are not based on any physical model. The main problem of those
classification criteria, which are based on the high-energy emission
alone, is that there is a significant overlap, which cannot be
quantified, between Collapsars and non-Collapsars in all of them. As
a result, the quality of the classification of any of these
phenomenological methods cannot be quantified and it is therefore
impossible to estimate the fraction of misclassified GRBs. This poses
a major problem in using such a method especially since the sample
of GRBs with ``good data" (afterglow detection, good localization,
redshift measurements etc.), is small and very sensitive to
misclassification. Other attempts used a statistical approach  and
tried to  evaluate the overlap between the two populations by
fitting the distribution of GRBs with two underlying distributions.
In this case two lognormal distributions \citep{Horvath02,Levesque10}.
The quality of such classification schemes depends entirely on
similarity of the true distribution of the two populations to the
arbitrarily chosen fitted distribution. Such approach can be trusted
only if we know, for example based on physical arguments, what is
the underling distribution of at least one of the two populations.
Recently, in \citet{B11a} we have shown, based on generic physical
properties of the Collapsar model, that  at short durations the
Collapsar distribution is flat. Namely the number of Collapsars per
unit duration at short durations is independent of the duration.
Building on this result we estimate here the probability that a GRB
with a given duration is a Collapsar or not. Not surprisingly this
probability depends on the detector and we calculate it for the
three  major GRB detectors: BATSE, \swift and \Fermi GBM.  An
improved version of this method is obtained by adding a hardness
dependence.  We present a refined probability distribution that is
based  on both the duration and the hardness. Needless to say our
method is statistical in nature. We cannot determine whether a
specific burst is a Collapsar or not, but we can give a probability
estimate for this question.

We begin, in section 2,  with a discussion of the Collapsars' duration
distribution and an analysis of the observed duration distribution
of GRBs . In section 3 we calculate the probability that an observed GRB is a
non-Collapsar. Our results imply that  short duration
Collapsars have been wrongly classified as non-Collapsars,
mostly in \swift sample, and this has lead to potential
misinterpretation of some of the observed data. In section 4 we discuss
the consistency  of our findings with some
of the recent studies and the implications on the inferred
properties of non-Collapsars. We summarize our results and their
implications in section 5.  We provide a table of the
probabilities for each one of the  observed \swift bursts with
$T_{90}< 2 $ sec to be a non-Collapsars in an Appendix.

\section{The observed GRB duration distribution}
\label{s.dNdT}

Within the  Collapsar model a GRB can only be produced  after the
jet has emerged from the surface of the collapsing star. We
\citep{B11a}  have recently shown that this leaves a distinctive
mark on the observed duration distribution: it is flat at durations
shorter than the typical breakout time of the jet from the star
(about a few dozen seconds modulo the redshift ). {In a nutshell,
this result arises from a simple fact. The burst duration is the
difference between two quantities: the engine operating time and the
jet breakout time. Under quite general conditions the resulting
distribution is flat at durations that are shorter than the typical
jet breakout time. Indeed, when we plot in fig. (\ref{fig.dNdT_all})
the quantity $dN_{GRB}/dT_{90}$ instead of the traditionally shown
 ${dN_{GRB}}/{d\log (T_{90}) }$  \citep[e.g. fig. \ref{fig.dNdlogT};][]{Kouveliotou93}   this flat distribution is evident.
 The plateau appears  over about an order of magnitude in duration around  a few seconds,
 in the GRB duration distributions of BATSE, \swift and \Fermi GBM, as
 depicted in fig. \ref{fig.dNdT_all}.
The duration is characterized by $T_{90}$ during which 90\% of the  fluence is
accumulated.

At the short end the distribution is rising towards shorter
durations. This ``bump" in the duration distribution is inconsistent
with a Collapsar origin for most of the short duration GRBs. This
simple conclusion is consistent with other evidence that a second,
non-Collapsar, population of short duration GRBs exists with a
different origin than the longer ones.
\citep[e.g.][]{Kouveliotou93,Barthelmy05,Fox05,Gehrels05,Nakar07}.

To quantify the non-Collapsars' duration distribution we make  joint fits to the
overall duration distributions, including the Collapsar distribution at durations longer than the plateau.
Although  we are interested  only in the short duration regime,
where the duration distribution of Collapsars
is flat, inclusion of the long end of the distribution is needed to determine the height of the plateau.
To test the robustness of our result we fitted various functional forms
for the distribution at long durations and verified that
the height of the plateaus are  consistent within the  errors.
The results presented here employ a plateau
below the typical observed breakout time, $T_B$, and a powerlaw with an
exponential cutoff above it.
For the non-Collapsars we find that the best fitted distribution function is a
lognormal.
Overall we fit the duration distributions  to the function:
\begin{equation}\label{eq.dNdT.GRB}
\frac{dN_{GRB}}{dT_{90}}=A_{NC}\frac{1}{T_{90}\sigma\sqrt{2\pi}}e^{-\frac{(\ln T_{90}-\mu)^2}{2\sigma^2}}+
A_C  \left\{
\begin{array}{lc}
1 & T_{90}\leq T_B \\
\left(\frac{T_{90}}{T_B}\right)^\alpha e^{-\beta(T_{90}-T_B)} & T_{90}>T_B,
\end{array}
\right.
\end{equation}
where the first term corresponds to non-Collapsars and the second one to Collapsars.

We consider the data sets of BATSE
\footnote{http://swift.gsfc.nasa.gov/docs/swift/archive/grb\_table,
from April 21, 1991 until August 17, 2000.},
\swift\footnote{
http://gammaray.msfc.nasa.gov/batse/grb/catalog/current/,
from  December 17, 2004 until February 20, 2012.},
and \Fermi GBM
\footnote{http://heasarc.gsfc.nasa.gov/W3Browse/fermi/fermigbrst.html,
from  July 17, 2008 until July 9, 2010.}.
We limit the data to the duration regime of 0-200 sec, which
is enough to obtain a good constraint of the plateau hight.
We verified that changing this range to 0-1000 sec has no
significant effect on our results. We fit each sample with a
distribution function according to eq. (\ref{eq.dNdT.GRB}). After
using the normalization that the integral of $dN_{GRB}/dT_{90}$ over
the duration range equals the number of observed GRBs, we are left
with seven free parameters. We obtain good fits with $\chi^2$ per
degrees of freedom (DOF) of 0.9, 1.3, 1.1 for the BATSE, \swift and
\Fermi GBM  respectively. The corresponding  parameters are given in
table \ref{table.Param_fit_all} and Fig. \ref{fig.dNdT_all} depicts
the resulting distribution functions and the data. We find
plateaus that extend up to $T_B\sim20$ sec in the  BATSE and \Fermi GBM
durations distributions, and up to $T_B\sim10$ sec in \swift. This
is consistent with our expectations from the Collapsar model
\citep{B11a}.

 \begin{figure}
\includegraphics[width=7in]{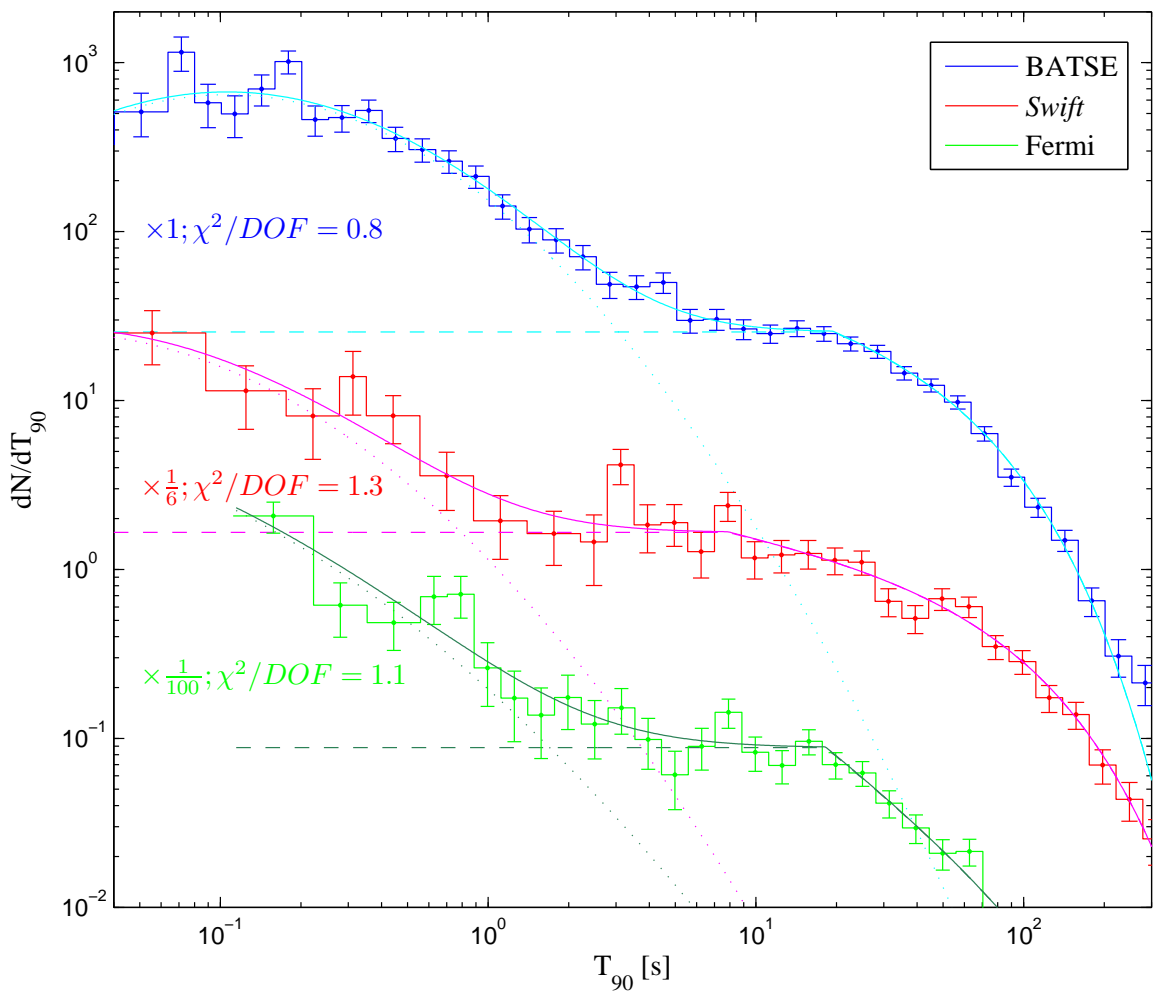}
  \caption{
  The $T_{90}$ distributions, $dN/dT_{90}$, of BATSE (red),
  {\it Swift} (blue) and Fermi GBM (green) GRBs, binned into equally spaced logarithmic bins.
  Bins with less than 5 events are merged with their neighbors to get more reliable
  statistical errors. Note that the quantity $dN/dT$  is depicted and
  not ${dN}/{d\log (T)}$ as traditionally shown in such plots \citep[e.g.,][]{Kouveliotou93}.
  The combined best fitted distribution functions of both Collapsars and non-Collapsars  are shown  with  solid
  lines. The dotted and dashed lines depict the distributions of non-Collapsars and Collapsars
  respectively in each data set.
   }
  \label{fig.dNdT_all}
\end{figure}

\begin{table}[h]
  \centering
  \caption{Best fit parameters}
  \label{table.Param_fit_all}
  \small{
 \begin{tabular} {c c c c c c c c}
Detector & $A_{NC}$ &  $\mu$ & $\sigma$ & $A_{C}$ & $T_B$ (s) & $\alpha$ & $\beta$\\
\hline
BATSE    & $545$ & $-0.5\pm0.1$ & $1.32\pm0.07$ & $25.5^{+1.9}_{-1.4}$ & $19.4^{+2.5}_{-4.2}$  & $-0.33\pm0.2$ & $0.019\pm0.003$\\
\swift   &$42$   & $-1.5\pm0.5$   & $1.5\pm0.4$            & $10.0\pm2.3$ & $7.9\pm3.5$           & $-0.3\pm0.2$ & $0.01\pm0.003$\\
\Fermi GBM& $128$& $-1.5\pm0.6$   & $1.9\pm0.4$            & $8.8\pm0.2$  & $18.2^{+1.3}_{-11.6}$ & $-1.2\pm0.36$  & $0.008\pm0.001$\\

\hline
\end{tabular}
}
\end{table}

\section{The non-Collapsars probability function}
\label{s.fNC}

The probability that a GRB with a given $T_{90}$ is a non-Collapsar is
given by the fraction of non-Collapsars within the observed GRBs at a
given duration:
\begin{equation}\label{eq:fNC}
f(T_{90})=A_{NC}\frac{1}{T_{90}\sigma\sqrt{2\pi}}e^{-\frac{(\ln T_{90}-\mu)^2}{2\sigma^2}}\left(\frac{dN_{GRB}}{dT_{90}}\right)^{-1},
\end{equation}
where $dN_{GRB}/dT_{90}$ is given by eq. (\ref{eq.dNdT.GRB}).
To estimate the errors in \f we simulate,
for each one of the samples,  distributions of  $T_{90}$ drawn randomly  from
the best fitted distribution function $dN_{GRB}/dT_{90}$.
We then bin the simulated data sets, and repeat the process of
parameter fitting using eq.(\ref{eq.dNdT.GRB}) to obtain \f. We
repeat this processes $1000$ times and look for the ranges of \f
that encompasses $68\%$ of the cases. Fig. \ref{fig.fNC_all} depicts
\f$(T_{90})$  for the BATSE, {\it Swift} and \Fermi GBM samples. The
solid lines depict \f, calculated from the  observed data, and the
blue region describes the $1 \sigma$ error estimate. Table
\ref{table.Param_fit_all}  lists the $T_{90}$ values that correspond
to some selected probabilities for the three detectors.

These results clearly show that the choice of $T_{90}=2$ sec as a threshold
to identify non-Collapsars is suitable for BATSE, and possibly also for \Fermi GBM.
A BATSE (Fermi GBM) burst with $T_{90} = 2 $ sec has a probability $>70\%$ ($\gtrsim40\%$) to be a non-Collapsar.
However, the probability of a similar {\it Swift} burst   to be a non-Collapsar
is only $0.16\pm0.14$. It is most likely a Collapsar!
The level of false identification, for a given duration threshold, $T_{th}$ can be seen in  Fig. \ref{fig.fC_tot_all} that depicts  the integrated fraction of Collapsars (out of the total number of GRBs) with duration $T_{90}\leqslant T_{th}$.
The total number of Collapsars with duration $T_{90}<2$ sec in the \swift sample
is estimated to be $19\pm5$ out of 53 GRBs. Thus,  an arbitrary \swift  sample selected with the
 $T_{th}=2$ sec criterion contains about than 40\% Collapsars that have been misclassified as non-Collapsars.
This should be  compared with  about 10\% and 15\% of misclassified
Collapsars in the corresponding BATSE and \Fermi samples (see fig.
\ref{fig.fC_tot_all}) with the same $T_{th}$. The criterion
$T_{th}=2$ sec for selecting non-Collapsars in \swift is simply
very bad for most studies of non-Collapsars.

Any single criteria that should distinguish according to the durations between Collapsars and non-Collapsars should be detector dependent.
A longer $T_{th}$  increases the size of the  SGRBs sample (that are supposedly
non-Collapsars) but it increases at the same time the number of misidentified Collapsars in the sample.  A shorter  threshold yields smaller but cleaner samples. The specific choice of $T_{th}$  should be considered for each study, balancing the need of a large sample with the importance of purity. A reasonable choice that should be adequate to many studies is  choosing  the threshold probability as \f$=0.5$. This reconciles between
these conflicting requirements and allows us to classify both Collapsars and non-Collapsars
with a single criterion. Adopting this probability we find that the corresponding $T_{90}$ threshold values are:
$T_{th}=0.8\pm0.3$ sec for \swift,
$T_{th}=1.7^{+0.4}_{-0.6}$ sec for \Fermi GBM  and
$T_{th}=3.1\pm0.5$ sec for  BATSE.
The total number of misclassified Collapsars in samples
selected according to these criteria constitute about 20\% of the  \swift
samples and $\sim14\%$ of the BATSE and Fermi GBM samples (fig. \ref{fig.fC_tot_all}).

  \begin{figure}
\includegraphics[width=7in]{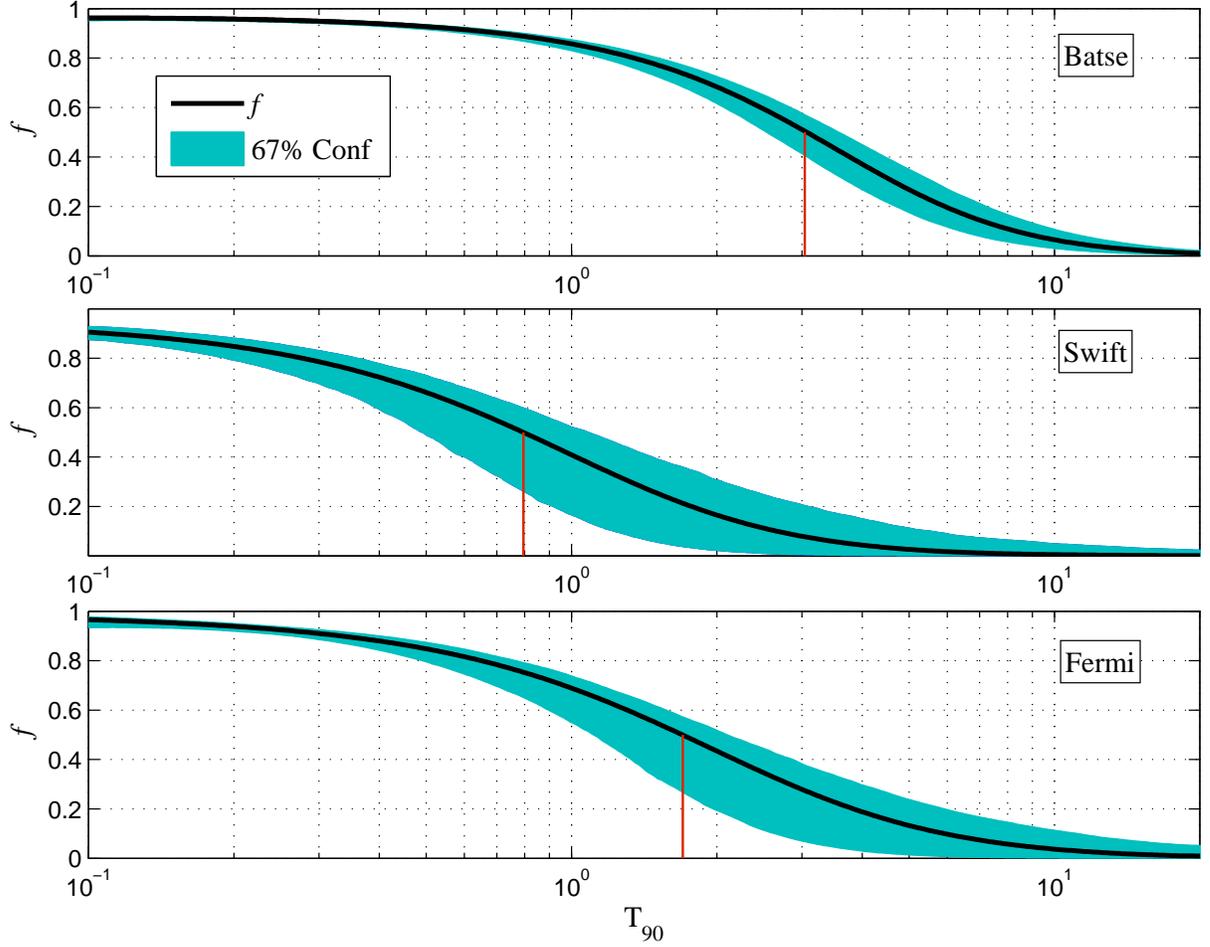}
  \caption{The fraction \f of non-Collapsars  as a function of $T_{90}$ for  BATSE, {\it Swift} \& Fermi GBM (from top to bottom).
  This fraction represents the probability that a GRB with an observed duration $T_{90}$ is a non-Collapsars.
   The shaded regions represent the 68\% confidence range.
   The red vertical lines mark the values of $T_{90}$ where \f$=0.5$ (See table \ref{table.fNC_all} for
   numeric values of $T_{90}$ that correspond to some selected \f values).}
  \label{fig.fNC_all}
\end{figure}

\begin{table}[h]
  \centering
  \caption{The $T_{90}$ (sec) that corresponds to some selected \f values in the three satellites}
  \label{table.fNC_all}
\begin{tabular} {c @{}c@{}| c c c c c c c}
 \hline\hline
          &\f & 0.9 & 0.8 & 0.7 & 0.6 & {\bf 0.5} & 0.4 & 0.3\\
 Satellite&  &     &     &     &     &     &     &    \\
  \hline
BATSE     &  &$0.7^{+0.1}_{-0.1}$  &$1.4^{+0.2}_{-0.2}$  &$1.9^{+0.3}_{-0.3}$  &
             $2.5^{+0.4}_{-0.4}$  &${\bf 3.1^{+0.5}_{-0.5}}$  &$3.8^{+0.7}_{-0.7}$  &$4.7^{+0.9}_{-0.9}$ \\

\swift    &  &$0.11^{+0.05}_{-0.11}$  &$0.3^{+0.1}_{-0.1}$  &$0.4^{+0.1}_{-0.1}$  &
             $0.6^{+0.2}_{-0.2}$  &${\bf 0.8^{+0.3}_{-0.3}}$  &$1.0^{+0.5}_{-0.4}$  &$1.3^{+0.7}_{-0.6}$ \\

\Fermi GBM&  &$0.3^{+0.1}_{-0.2}$  &$0.7^{+0.1}_{-0.2}$  &$1.0^{+0.2}_{-0.3}$  &
            $1.3^{+0.3}_{-0.4}$  &${\bf 1.7^{+0.4}_{-0.6}}$  &$2.2^{+0.7}_{-0.9}$  &$2.8^{+1.1}_{-1.2}$  \\

\hline
 \end{tabular}
\end{table}

  \begin{figure}
\includegraphics[width=7in]{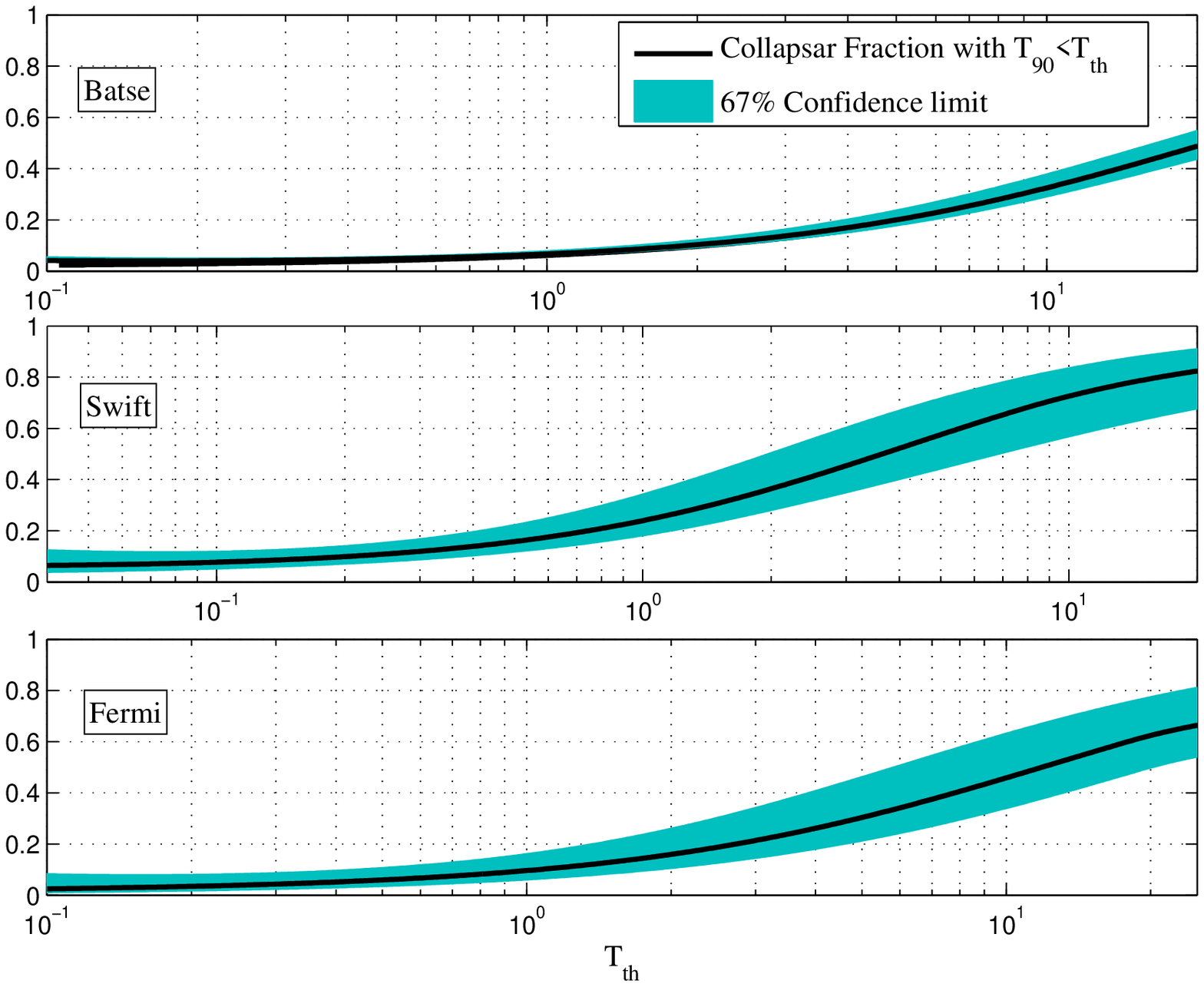}
  \caption{The integrated fraction  of Collapsars with duration $T_{90}\leqslant T_{th}$,
   for BATSE, {\it Swift} \& Fermi GBM  (from top to bottom). This is the fraction of {\bf Collapsars} that
   are misclassified as non-Collapsars with durations shorter that the threshold, $T_{th}$.
  The shaded regions represent the 68\% confidence limits.}
  \label{fig.fC_tot_all}
\end{figure}

\subsection{The non-Collapsar probability as a function of duration and hardness}\label{s.fNC_hardness}

As already mentioned short GRBs are harder on average than long ones \citep{Kouveliotou93}.
It is natural to expect that the ratio of non-Collapsars to
Collapsars and the probability function, \f, should increase with
the GRB hardness. Therefore the combination of duration and hardness
provides a stronger way to distinguish between  non-Collapsars and
Collapsars. To examine the duration-hardness probability we divide
the  samples into three hardness subgroups: soft, intermediate and
hard, and preform the same analysis on each subgroups: We fit a
duration distribution function  and calculate the probability
function, \f, and its $1\sigma$ variance. We consider two hardness
thresholds: The soft and intermediate subgroups are separated by the
average hardness of Collapsars which we estimate using GRBs with
$T_{90}>20$ sec, where  the contribution of non-Collapsars in all
samples is negligible. The intermediate and hard subgroups are
separated by the average hardness of non-Collapsars which is
estimated using GRBs with $T_{90}<0.5$ sec.
The spectral hardness of different satellite samples is quantified
differently for each detector, depending on the available
information for each database. For BATSE we use the hardness ratio
parameter, HR$_{32}$, defined as the ratio between the photon counts
in energy channel 3 (100 - 300 keV) and energy channel 2 (50 - 100
keV). In \swift and \Fermi GBM samples we use the powerlaw index
(PL) of the observed spectrum obtained by fitting a single powerlaw
in the energy range $15-150$ keV (\swift) or $10-2000$ keV (Fermi).
Note that in the \swift sample, only $\sim87\%$ of the GRBs have a
spectral fit to a single power-law. The spectrum of the other $13\%$
is fitted with a powerlaw+exponential cutoff, and we omit these
bursts from this analysis.

\begin{figure}[h]
\begin{center}$
\begin{array}{c}
\begin{array}{cc}
\hspace{-10mm}
\includegraphics[width=3.2in]{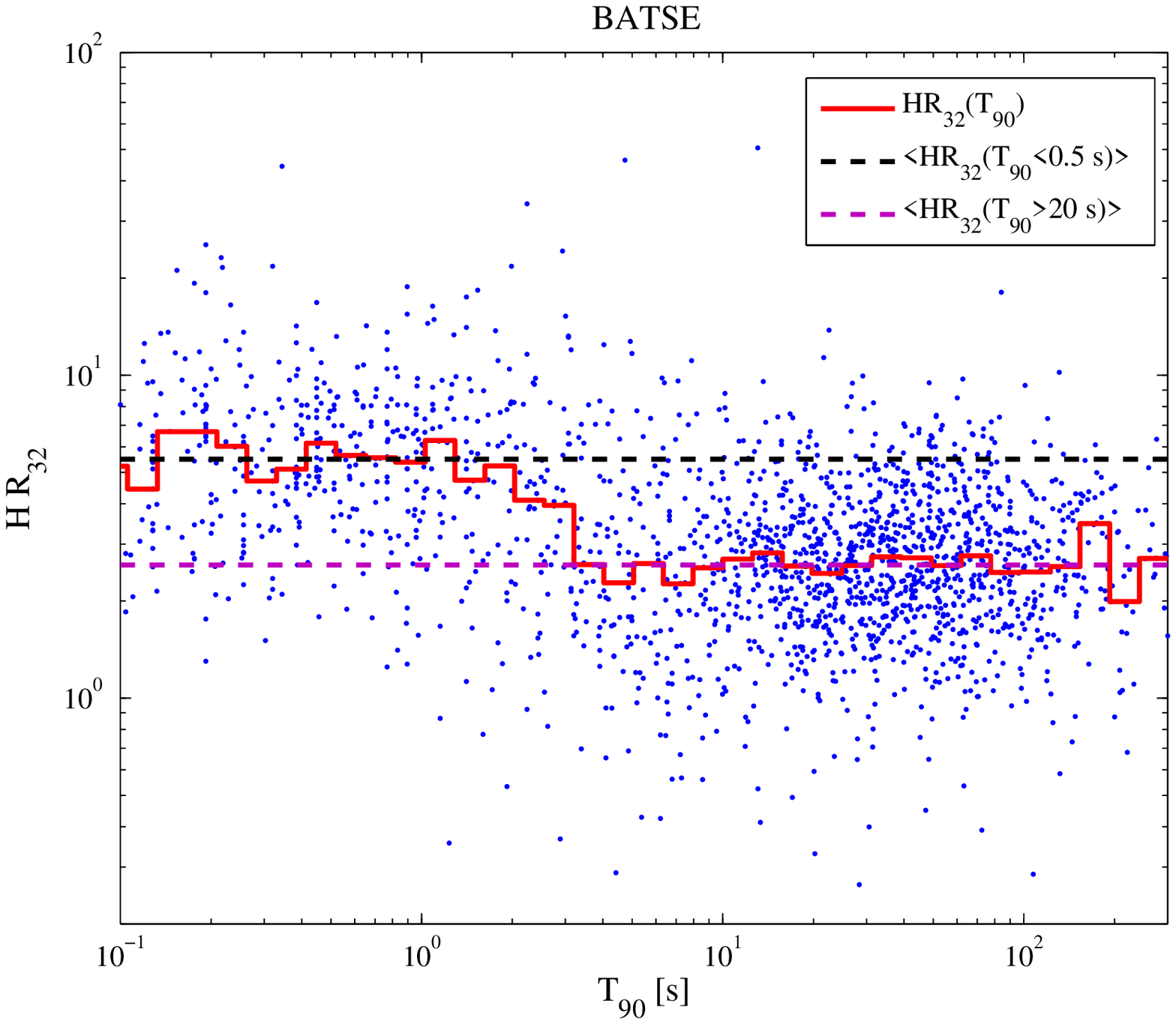} &
\hspace{-10mm}
\includegraphics[width=3.2in]{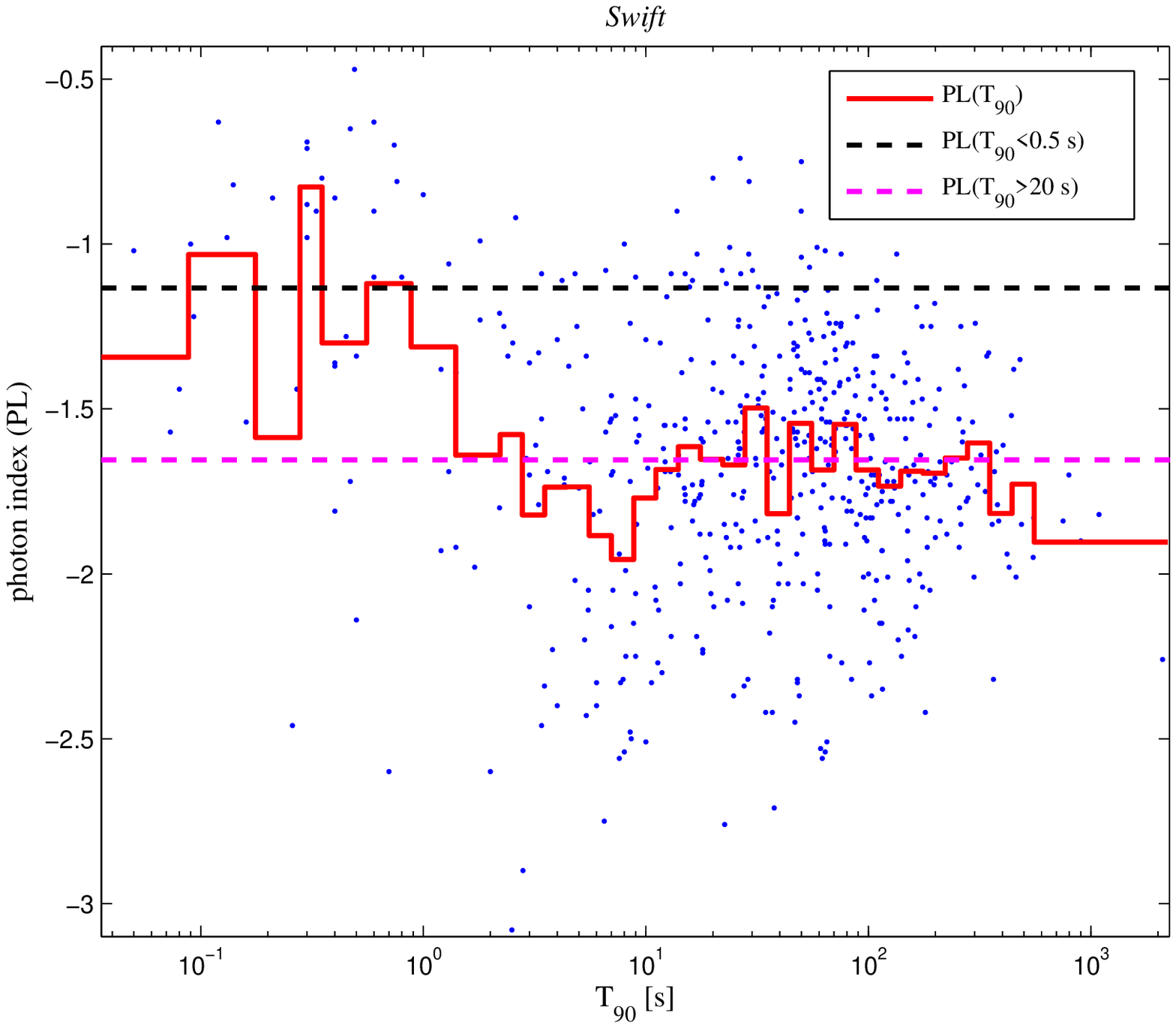}
\end{array}
\\
\hspace{-10mm}
\includegraphics[width=3.2in]{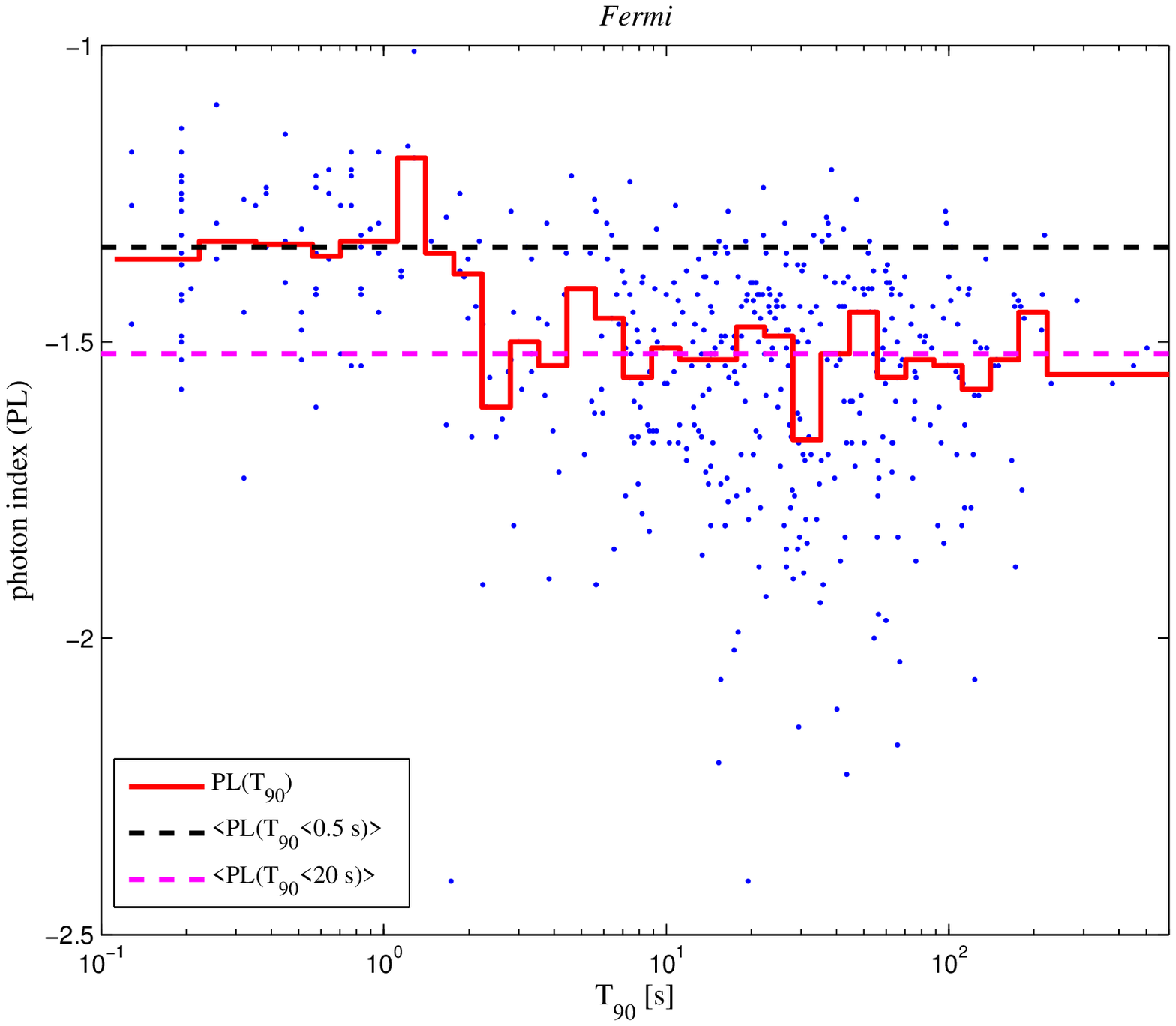}
\end{array}$
\end{center}
\caption{The hardness ratio of BATSE GRBs (top left)
and the powerlaw index of \swift GRBs (top right) and
\Fermi GRBs (bottom) as a function of $T_{90}$.}
\label{fig.hardness}
\end{figure}

If the distribution functions of Collapsars and non-Collapsars
don't depend strongly on the spectral hardness,
then varying the hardness threshold would only change the relative ratio of Collapsars to
non-Collapsars, while the overall shape of the duration distribution functions
of the two populations would remain unchanged.
To examine this we fit each hardness subgroup with the same distribution function as
in the full sample of the corresponding detector. We only rescale $A_{NC}$ and $A_{C}$
according to the number of non-Collapsars and Collapsars in the subgroup relative
to their number in the full sample.
We evaluate the ratio of non-Collapsars by dividing the number of GRBs with
$T_{90}<0.5$ sec in the subgroup with their number in the full sample.
The ratio of Collapsars is evaluated in a similar way using GRBs with
$T_{90}>20$ sec.
We find  good fits in all  hardness subgroups with $\chi^2/dof$ of $\sim 0.8-1.8$.
This good fit indicates  a weak  dependency of the duration distributions of Collapsars and non-Collapsars on the hardness.

Figure \ref{fig.dNdT_hardness} depicts the observed duration
distributions of the three hardness subgroups of each detector. The
harder subgroups have more prominent 'bumps' at short durations
together with relatively lower plateaus that become visible only at
longer durations. This is the expected behavior if the fraction of
non-Collapsars increases with the GRB hardness.
The  $dN/dT_{90}$ distributions and their $\chi^2/DOF$ values are shown in fig. \ref{fig.dNdT_hardness}.
The resulting probability functions of each subgroup are shown in figs.
\ref{fig.fNC_hardness_BATSE}-\ref{fig.fNC_hardness_Fermi}.
In table \ref{table.fNC_hardness}
we list the $T_{90}$ values that correspond to specific values of \f in each
subgroup.
In the hard BATSE subgroup non-Collapsars dominate
the distributions up to $T_{90}=7.8^{+1.4}_{-1.0}$ sec, while in the hard
\swift and \Fermi GBM samples non-Collapsars  dominate  up to $T_{90}=2.8^{+1.5}_{-1.0}$
and $T_{90}=5.4^{+3.9}_{-2.0}$ sec respectively.
The transition between Collapsars and non-Collapsars
in the intermediate subgroups roughly
follow the same values as in the complete samples:
$T_{90}=2.9\pm0.4$, $T_{90}=0.6^{+0.2}_{-0.3}$ and $T_{90}=1.6^{+0.8}_{-0.6}$
sec for BATSE, \swift and \Fermi GBM respectively.
In the soft subgroups non-Collapsars dominate only up to
$T_{90}=1.1^{+0.2}_{-0.3}$ sec in BATSE, $T_{90}=0.3^{+0.4}_{-0.2}$ sec
in \swift, and up to $T_{90}=0.6^{+0.6}_{-0.3}$ sec in \Fermi GBM.

\newpage

\begin{figure}[h]
\begin{center}$
\begin{array}{c}
\begin{array}{cc}
\hspace{-10mm}
\includegraphics[width=3.5in]{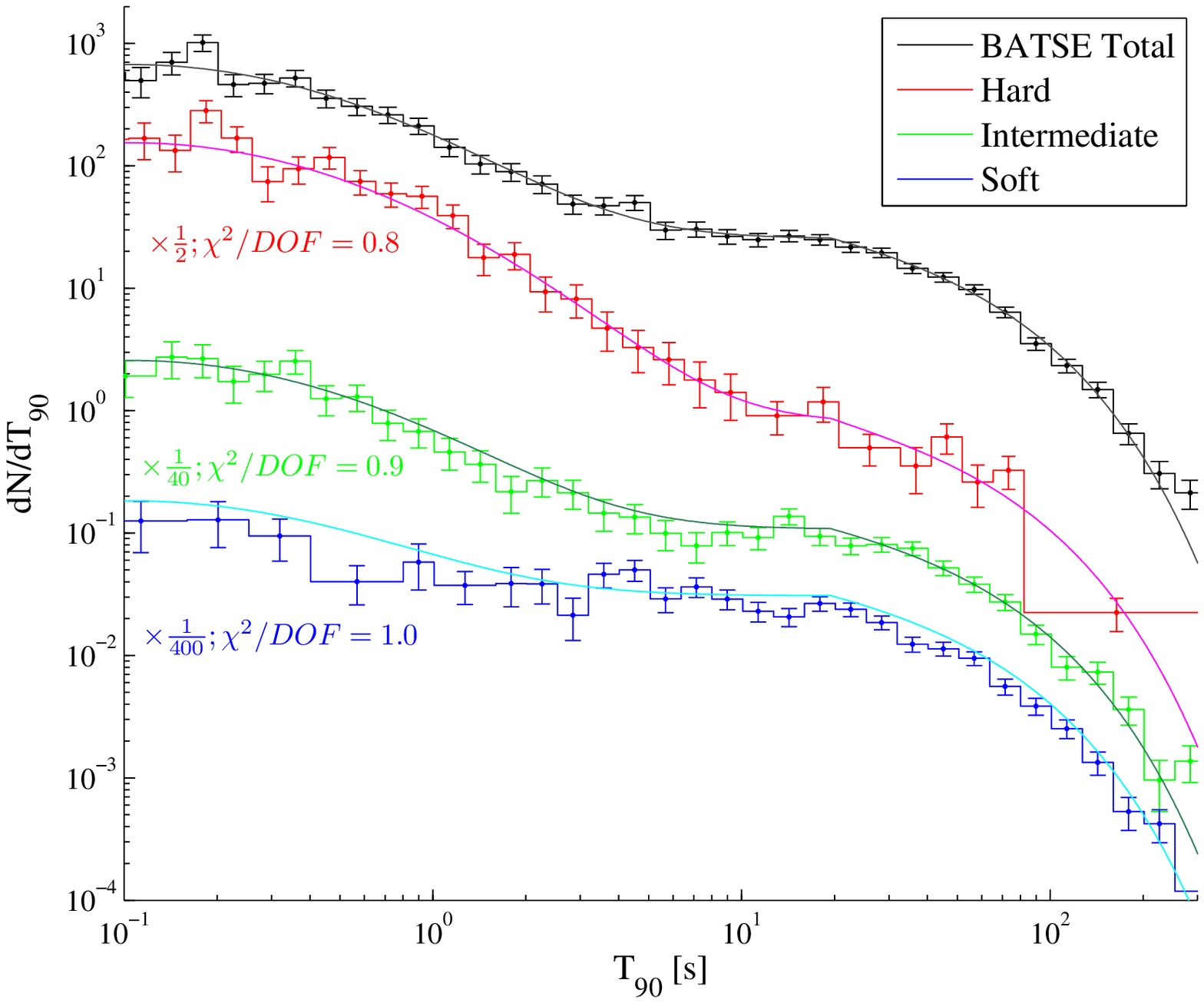} &
\hspace{-10mm}
\includegraphics[width=3.5in]{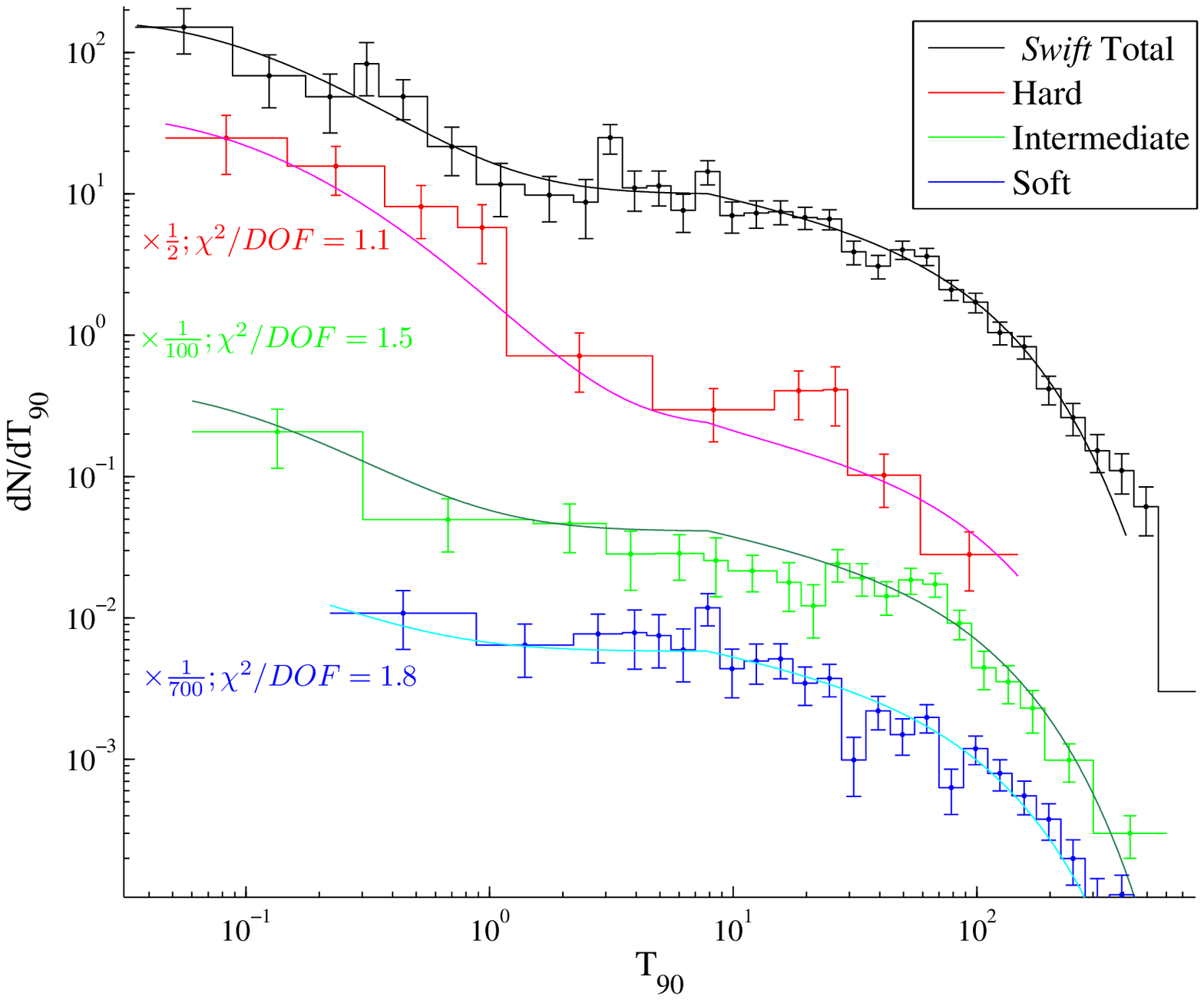}
\end{array}
\\
\hspace{-10mm}
\includegraphics[width=3.5in]{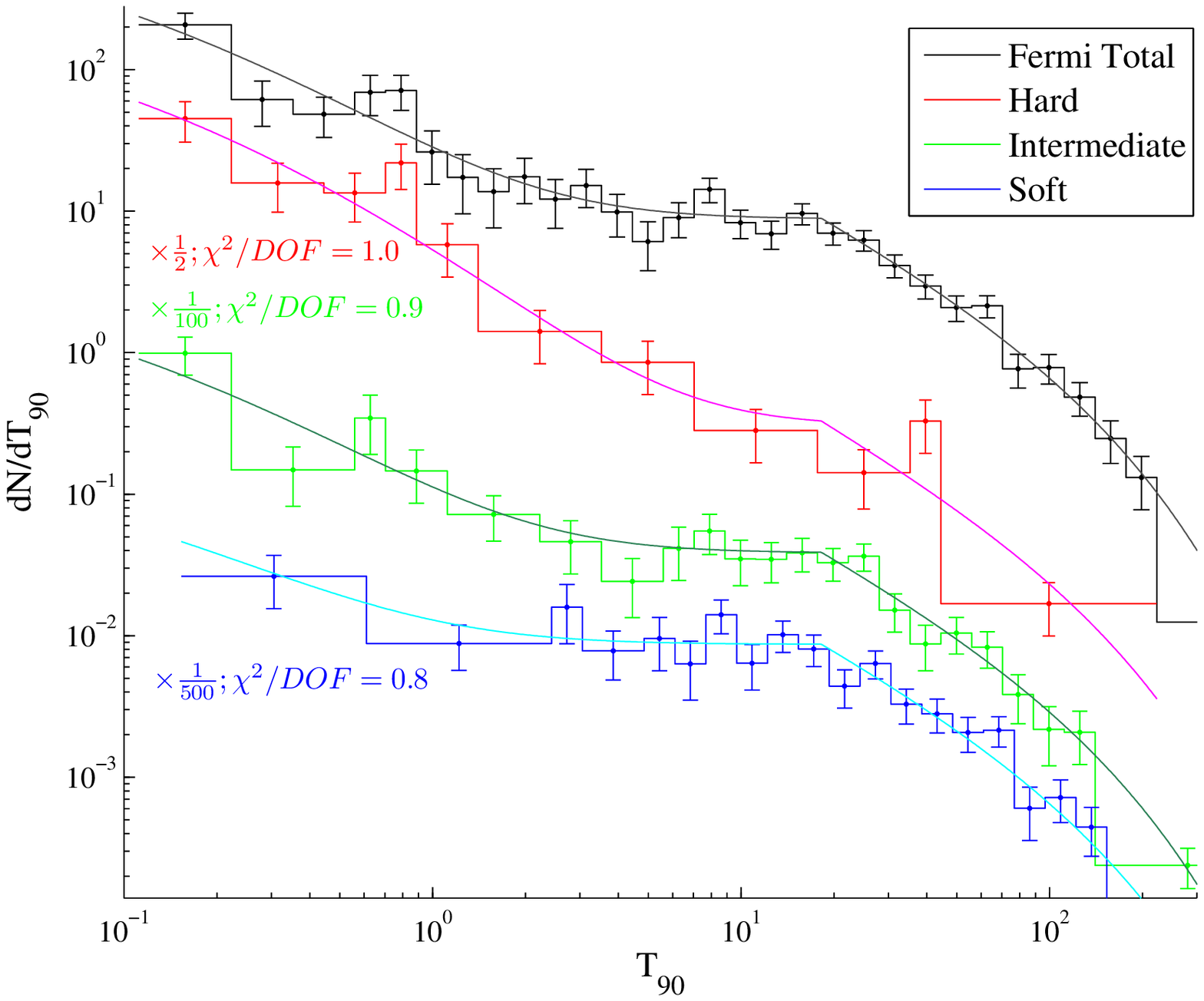}
\end{array}$
\end{center}
\caption{$dN/dT_{90}$, of the three hardness subgroups  of the BATSE (upper left)
  {\it Swift} (upper right) and Fermi GBM (lower) samples, binned into equally spaced logarithmic bins.
  Bins with less than 5 events are merged with their neighbors to reduce statistical errors.
  The best fitted joined distribution functions
  are marked  with  solid lines.
  }
  \label{fig.dNdT_hardness}
\end{figure}

  \begin{figure}
\includegraphics[width=5in]{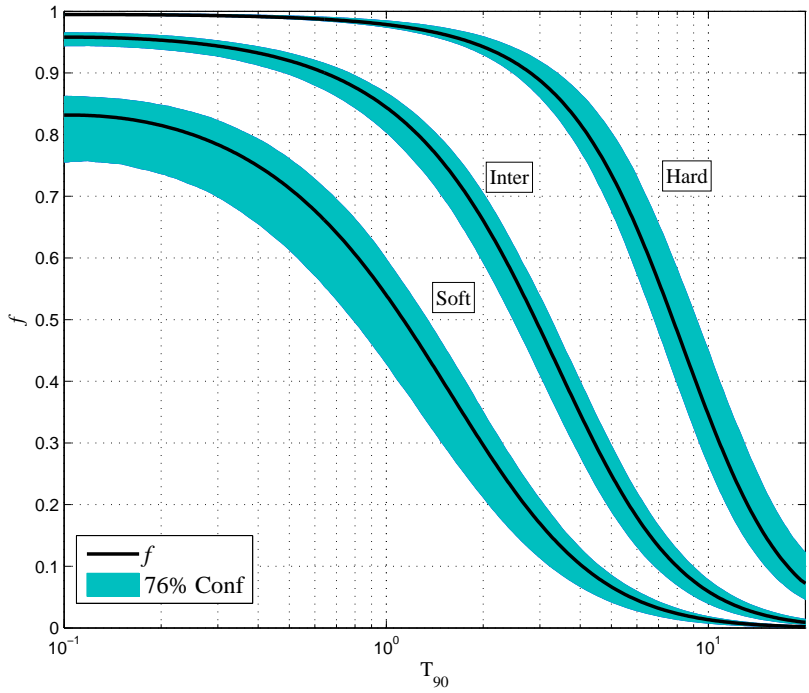}
  \caption{The fraction, \f, of non-Collapsars  as a function of the observed duration, $T_{90}$,
  in the 3 hardness subgroups of BATSE.}
  \label{fig.fNC_hardness_BATSE}
\end{figure}

  \begin{figure}
\includegraphics[width=5in]{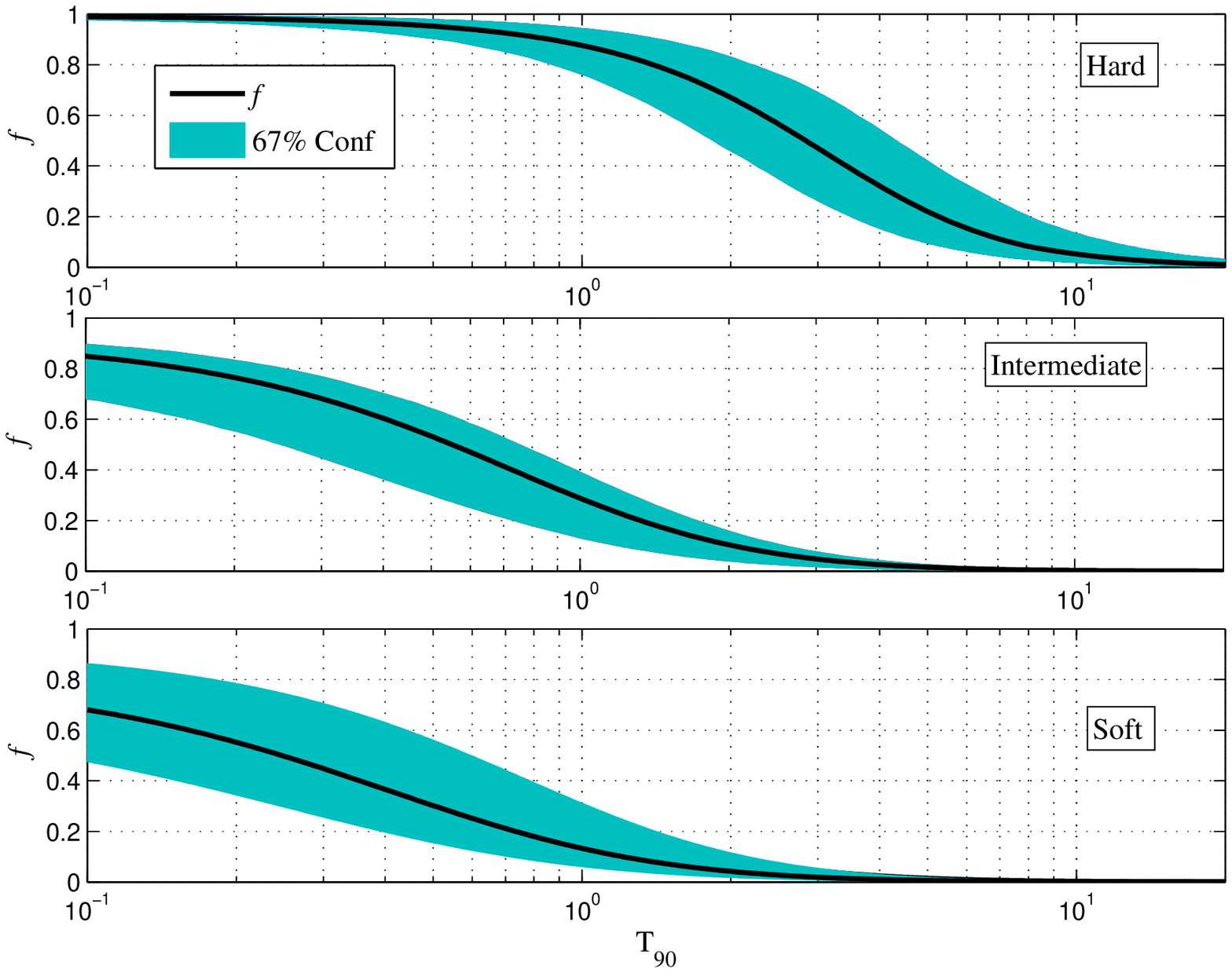}
  \caption{Same as fig. \ref{fig.fNC_hardness_BATSE} for \swift}
  \label{fig.fNC_hardness_Swift}
\end{figure}

  \begin{figure}
\includegraphics[width=5in]{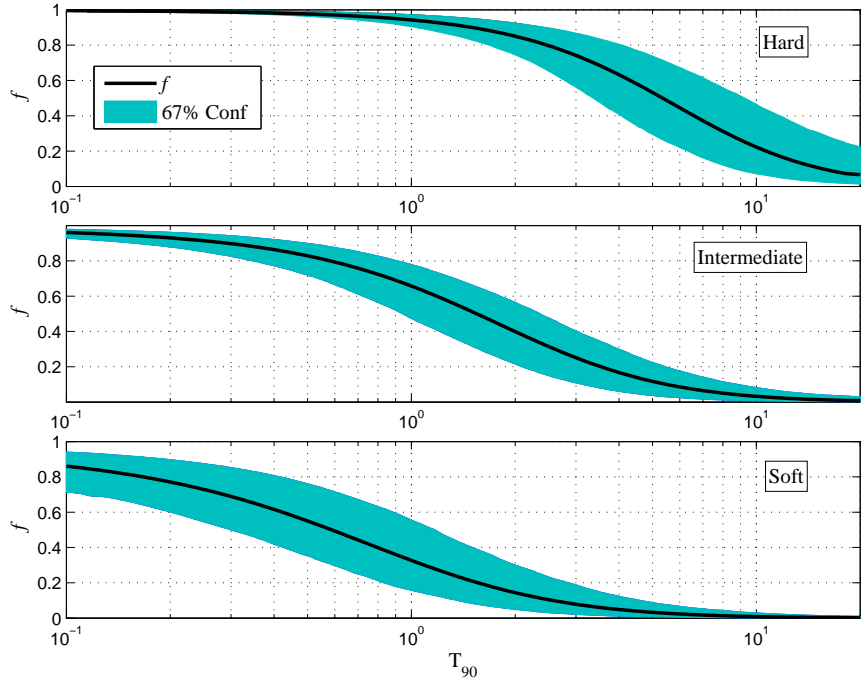}
  \caption{Same as fig. \ref{fig.fNC_hardness_BATSE} for \Fermi GBM}
  \label{fig.fNC_hardness_Fermi}
\end{figure}

\begin{table}[h]
  \centering
  \caption{The $T_{90}$ (s) at different \f in the three hardness subgroups for each satellite.}\label{table.fNC_hardness}

\begin{tabular} {@{} c |@{} c @{}| c c c c c c c|}
 \cline{2-9}
             &\diagbox[width=7em, height=3em, trim=lr]{Hardness}{\f}         & 0.9 & 0.8 & 0.7 & 0.6 & {\bf 0.5} & 0.4 & 0.3\\

  \hline

\multirow{6}{*}{\rotatebox{90}{\bf BATSE}} &
\shortstack[c]{ Hard \\ \\ {\scriptsize  (HR$_{32}>5.5$)}}&
$2.8^{+0.6}_{-0.3}$ & $4.2^{+0.8}_{-0.5}$ & $5.4^{+1.0}_{-0.6}$ &
$6.6^{+1.2}_{-0.8}$ & ${\bf 7.8^{+1.4}_{-1.0}}$ & $9.1^{+1.7}_{-1.2}$ &$10.8^{+2.1}_{-1.4}$\\

& & & & & & & &\\
& \shortstack{Intermediate \\ \\{\scriptsize ($5.5>$HR$_{32}>2.6$)}} &
$0.6^{+0.1}_{-0.2}$  &$1.3^{+0.2}_{-0.2}$  &$1.8^{+0.2}_{-0.3}$  &
$2.3^{+0.3}_{-0.4}$  &${\bf 2.9^{+0.4}_{-0.4}}$  &$3.6^{+0.4}_{-0.5}$  &$4.4^{+0.5}_{-0.6}$\\

& & & & & & & &\\
& \shortstack{Soft \\ \\  {\scriptsize ($2.6>$HR$_{32}$)}}&
...                     &$0.2^{+0.2}_{-0.2}$  &$0.5^{+0.2}_{-0.3}$  &
$0.8^{+0.2}_{-0.3}$  &${\bf 1.1^{+0.2}_{-0.3}}$  &$1.5^{+0.3}_{-0.4}$  &$2.0^{+0.3}_{-0.5}$ \\

\hline

\multirow{6}{*}{\rotatebox{90}{\bf \swift}} &

\shortstack[c]{ Hard \\ \\ {\scriptsize  (PL$>-1.13$)}}&
$0.9^{+0.6}_{-0.4}$  &$1.4^{+0.9}_{-0.5}$  &$1.9^{+1.1}_{-0.6}$  &
$2.3^{+1.3}_{-0.8}$  &${\bf 2.8^{+1.5}_{-1.0}}$  &$3.4^{+1.9}_{-1.1}$  &$4.2^{+2.2}_{-1.4}$  \\

& & & & & & & &\\
& \shortstack{Intermediate \\ \\{\scriptsize ($-1.13>$PL$>-1.65$)}} &
$0^{+0.09}$               &$0.16^{+0.09}_{-0.16}$  &$0.3^{+0.1}_{-0.2}$  &
$0.4^{+0.2}_{-0.2}$  &${\bf 0.6^{+0.2}_{-0.3}}$  &$0.7^{+0.2}_{-0.4}$  &$1.0^{+0.3}_{-0.5}$  \\


& & & & & & & &\\
& \shortstack{Soft \\ \\  {\scriptsize ($-1.65>$PL)}}&
$0^{+0.05}$             &$0^{+0.18}$              &$0.09^{+0.22}_{-0.09}$  &
$0.16^{+0.28}_{-0.16}$  &${\bf 0.3^{+0.4}_{-0.2}}$  &$0.4^{+0.4}_{-0.2}$  &$0.5^{+0.5}_{-0.3}$  \\


\hline

\multirow{6}{*}{\rotatebox{90}{\bf \Fermi GBM}} &

\shortstack[c]{ Hard \\ \\ {\scriptsize  (PL$>-1.34$)}}&
$1.5^{+1.0}_{-0.5}$  &$2.5^{+1.6}_{-0.8}$  &$3.4^{+2.2}_{-1.1}$  &
$4.3^{+3.0}_{-1.5}$  &${\bf 5.4^{+3.9}_{-2.0}}$  &$6.6^{+5.1}_{-2.5}$  &$8.2^{+7.1}_{-3.3}$  \\

& & & & & & & &\\
& \shortstack{Intermediate \\ \\{\scriptsize ($-1.32>$PL$>-1.52$)}} &
$0.3^{+0.2}_{-0.2}$  &$0.6^{+0.3}_{-0.2}$  &$0.9^{+0.5}_{-0.3}$  &
$1.2^{+0.6}_{-0.5}$  &${\bf 1.6^{+0.8}_{-0.6}}$  &$2.0^{+1.0}_{-0.8}$  &$2.6^{+1.4}_{-1.0}$  \\

& & & & & & & &\\
& \shortstack{Soft \\ \\  {\scriptsize ($-1.52>$PL)}}&
$0.06^{+0.14}_{-0.06}$  &$0.17^{+0.25}_{-0.17}$  &$0.3^{+0.4}_{-0.2}$  &
$0.4^{+0.5}_{-0.2}$  &${\bf 0.6^{+0.6}_{-0.3}}$  &$0.8^{+0.7}_{-0.4}$  &$1.1^{+0.9}_{-0.5}$  \\

\hline\hline
 \end{tabular}
\end{table}


The probability functions we obtained  here can be used to classify the GRBs detected
by BATSE, \swift and \Fermi GBM  according to their duration and hardness.
For GRBs that cannot be assigned to one of those hardness subgroups
(e.g the 13\% of \swift GRBs whose spectra are fitted with a powerlaw+exponential cutoff) the
classification can be done using the overall probability function of the complete \swift sample.
For  GRBs that have been detected by HETE and Integral  we
recall that the spectral window observed by HETE is $8-500$ keV,
which  is closer to the \swift/BAT while  Integral, on the other
hand, observes at a spectral range of 15 keV - 10 MeV, which is
closer to BATSE's range. As a first approximation one can use the
corresponding \f values of these detectors.

In Appendix \ref{App:SwiftSGRBTable} we collect all \swift GRBs
with  $T_{90}<2$ observed to date. The table includes also  GRBs
with a hard short spike plus a soft extended emission and a number of
other GRBs with duration $>2$ sec that are sometimes considered as
possible non-Collapsars. For each GRB we calculate the probability
to be a non-Collapsar from the duration
and power-law index ,\f$(T_{90},PL)$.
For those GRBs with a spectral fit of a powerlaw+exponential
cutoff, we  calculate the probability to be a non-Collapsar from
the duration alone, \f$(T_{90})$. Important GRBs are
emphasized with bold text. The table also includes a few important
GRBs detected by HETE or Integral. For these GRBs we estimate
\f$(T_{90})$ using the probability functions of \swift and BATSE
respectively. This table can be used to evaluate the contamination
by Collapsars in present samples of SGRBs and to select  low
contamination samples in future studies.

\section {Consistency checks with studies of contaminated samples}

The commonly used criterion to distinguish Collapsars from
non-Collapsars is the duration ($T_{90}\gtrless2$ sec). This
criterion is applied to GRBs that are detected by all $\gamma$-ray
satellites including \swift, that supplies the largest number of
well localized short duration GRBs. As we have shown earlier \swift
GRBs with $T_{90}>0.8$ sec have high probability to be Collapsars.
This could have  led to a ``Collapsar contamination" in current
\swift samples of SGRBs that are based on the 2 sec criterion, and
might have affected the results of studies based on these samples.
Interestingly such studies
\citep[e.g.][]{Berger09,Berger11}, have shown that the environments of SGRBs
are different than the environments of LGRBs. Whereas LGRBs are
associated with intensive star formation, arise in low metallically
irregular star forming galaxies \citep[see however][for
examples of high metalicity LGRB
hosts]{Levesque10b,Levesque10c,Savaglio12} and are concentrated
towards star forming regions in their galaxies. SGRBs are associated
with a broad distribution of galaxy types and arise in hosts with a
broad range of star formation rate and metallicities and show a
larger scatter in the distance distribution from their hosts'
centers. One may wonder how these results are consistent with our
claim that the 2 sec classification is not valid for the \swift
sample.

Table \ref{table_Berger} lists  the GRBs and the host galaxy
characteristics used in the \citet{Berger09,Berger11} sample. It
also includes the probability that the associated SGRB are
non-Collapsars (based the combination of duration and power-law
index). In addition to eight `classically selected' SGRBs
($T_{90}<2$ sec) this sample includes also four GRBs with $T_{90}>2$
sec. These GRBs are characterized by a short hard initial spike
followed by a long tail of softer emission. They are often
considered as non-Collapsars since their initial spikes resemble a
classical SGRB \citep[see][]{Nakar07}. However, since the overall
duration is not well defined our classification scheme cannot
attribute a non-Collapsar probability to these bursts.

Even though our probabilistic approach is incapable of determining
whether a specific burst is a Collapsar or not, a clear picture
emerges from table \ref{table_Berger}. Four out of eight
classifiable bursts are non-Collapsars at very high probabilities.
Two bursts are almost certainly Collapsars while the last two are
marginal: the probability of each one of those two to be a Collapsar
is larger than 60\%, however the probability that both are Collapsars is
less than 50\%. These fractions are consistent  with what is
expected, according to our analysis, from a \swift sample with a 2
sec criteria for which $\sim 60\%$ of the bursts should be
non-Collapsars and the rest Collapsars.

Within the sub-sample of four non-Collapsars we observe a large
spread in SFRs, in specific SFRs and in galactic luminosities.
Distances from the center of the host have  typically large
observational error, but at least one is quite far from the
center  ($\sim44^{+12}_{-23}$ kpc). There is not enough data to determine the metallicity. These
results show a large spread in the observed quantities, in a large
contrast with the rather narrowly distributed host properties of
LGRBs (Collapsars). This is similar to the conclusion of
\citet{Berger09,Berger11}. It demonstrates that also when a less
contaminated, but smaller, sample is examined non-Collapsars hosts
have a different distribution than Collapsar hosts and consequently
that the two populations have different progenitors. On the other
hand, as expected,  the properties of the hosts of the two Collapsar
candidates are fully consistent with those of typical LGRB hosts.
Finally, the properties of the hosts of the two bursts with
marginal classification are also consistent with being either
Collapsar or non-Collapsar hosts.

The conclusion that the properties of the non-Collapsars' hosts
are widely distributed whereas those of the Collapsars' hosts are
narrowly distributed implies that our classification is consistent
with the results of \citet{Berger09,Berger11} even though the latter are
based of a significantly contaminated sample. A wide distribution
contaminated by  a narrowly distributed population retains it basic
feature of a wide distribution, and this is what happens here. The
non-Collapsars within the \citet{Berger09,Berger11} SGRB sample are
numerous enough to result in a wide distribution that is
significantly different from the one of Collapsars. However, while
our conclusions are in line with the basic results of
\citet{Berger09,Berger11}, the details of the distribution,  such as
the ratio of high SFR to low SFR hosts or the distribution of
distances from center, are influenced by the contamination and a
quantitative study of the distribution of the host properties should
take this factor into account.

The possible effects of contaminating Collapsars on studies of
properties of SGRBs vary from one study to another. Different
samples have different contaminations and different properties are
influenced differently. The probabilities given in  appendix A can
be used to evaluate the likelihood that different bursts are
non-Collapsars or Collapsars and with these to estimate the quality
of a specific sample and the significance of results based on this sample. In general
one should proceed with care before adopting simply the results of a
study of an SGRB sample as reflecting the properties of
non-Collapsars. In this context it is interesting to mention a few
GRBs that play a major role in the current view of non-Collapsar
properties. GRB 060121 and GRB 090426 are two SGRBs with a secure
host at redshift $\geqslant 2$ that have led to the suggestion of a
high redshift non-Collapsar population. GRB 100424A has a redshift
of $z=1.288$. All other SGRBs with secure redshift are at $z
\leqslant 1$. We find that the probabilities that these bursts are
non-Collapsars are $0.17^{+0.14}_{-0.15}$, $0.10^{+0.15}_{-0.06}$
and $0.08^{+0.12}_{-0.04}$ for GRBs 060121, 090426 and 100424A
respectively. Surely, one cannot establish a new population of high
redshift non-Collapsars based on these events. Another pivotal burst
is 051221A; the only SGRB to date with a clear simultaneous
optical/X-ray break in its afterglow, which is used to measure its
beaming \citep{Soderberg06b,Burrows06}. We find that the probability
that GRB 051221A is a non-Collapsar is $0.18^{+0.08}_{-0.11}$. This
highlights our ignorance of the collimation (if there is any) of
non-Collapsar outflows. It also highlights the fact that no firm
conclusion can be drawn on non-Collapsars based on a single burst
that is classified using its high energy emission properties
alone.

\begin{table}[h]
  \caption{The sample of \citet{Berger09,Berger11} SGRBs.}
  \label{table_Berger}
  \small{
\begin{tabular} {c | @{} c @{} c c c c c@{} @{}c @{}c  c}
 \hline
 GRB    &   $T_{90}$  &   PL            &  \f$^a$        &$L_b$ &   SFR      & SFR/$L_b$   &{\sm 12+log(O/H)$^*$}&offset&ref\\
        &      {\sm(s)}      &                 &                &{\sm($L_*$)} &{\sm($M_\odot/yr$)}&{\sm($M_\odot/yr\cdot L_*$)} & &{\sm(kpc)}&\\
 \hline\hline
050709$^b$  & 0.07    &                 &  $0.92^{+0.02}_{-0.03}$    &  0.1  & 0.2     &  2      & 8.5&3.8& 1,2\\
061217      &  0.210  &  $0.86\pm0.30$  &  $1^{+0.00}_{-0.21}$    & 0.4  & 2.5     & 6.25    &    &$0-30$& 1,3\\
050509B     &  0.073  &  $1.57\pm0.38$  &  $0.87^{+0.04}_{-0.16}$ & 5    & $<0.1$  & $<0.02$ &    &$44^{+12}_{-23}$&1,2\\
060801      &  0.490  &  $0.47\pm0.24$  &  $0.95^{+0.03}_{-0.05}$ & 0.6  &  6.1    & 10.17   &    &$19\pm16$&1,3\\
\hdashline
070724A     &  0.400  &  $1.81\pm0.33$  &  $0.37^{+0.26}_{-0.17}$ & 1.4  & 2.5     & 1.79    &8.9 &$4.8\pm0.1$&1,4\\
070429B     &  0.470  &  $1.72\pm0.23$  &  $0.32^{+0.26}_{-0.15}$ & 0.6  &1.1      & 1.83    &    &$40^{+48}_{-40}$&1,3\\
051221A     &  1.400  &  $1.39\pm0.06$  &  $0.18^{+0.08}_{-0.11}$ & 0.3  & 1       & 3.33    & 8.2 or 8.7&$0.8\pm0.3$&1,5\\
060121$^b$  & 1.97    &                 &  $0.17^{+0.14}_{-0.15}$    &      &  1      &         &    &$\sim1$&1,3\\
050724      &  3(96)$^\dag$     &       &                         &   1  & $<0.05$ & $<0.05$ &    &2.6& 1,2\\
061006      &  $~0.5(123)^\dag$  &       &                        & 0.1 &  0.2     & 2       & 8.6&$1.3$&1,3\\
061210      &   0.2(85)$^\dag$  &      &                         &  0.9 & 1.2      & 1.33    & 8.8&$11\pm10$&1,3\\
070714B     &  3(64)$^\dag$  &          &                         & 0.1 & 0.4      & 4       &    &$\lesssim4$&1,3\\
\hline
 \multicolumn{9}{l}{\footnotesize $^a$ \swift GRBs with a single power-law spectral fit are assigned a probability \f$(T_{90},PL)$.}\\
 \multicolumn{9}{l}{\footnotesize \hspace{6pt}  Other GRBs can only be assigned a probability \f$(T_{90})$.} \\
 \multicolumn{9}{l}{\footnotesize $^b$ A GRB detected by HETE, \f$(T_{90})$ is estimated using \swift probability function.}\\
 \multicolumn{9}{l}{\footnotesize $\dag$  GRB with an extended softer emission}\\
 \multicolumn{9}{l}{\footnotesize $^*$ The metallicity is measured by the ratio of Oxygen to Hydrogen lines. The range of values}\\
 \multicolumn{9}{l}{\footnotesize \hspace{9pt}of $8.2-8.9$ shown in the table corresponds to $\sim0.3-1.6 Z_\odot$.}\\
 \multicolumn{9}{l}{\footnotesize References: 1) \citet{Berger09}; 2) \citet{Fox05}; 3) \citet{Fong10};}\\
 \multicolumn{9}{l}{\footnotesize  4) \citet{Berger09b}; 5) \citet{Soderberg06b}}\\
 \end{tabular}
 }
\end{table}

\section{Summary}

GRBs are widely classified as long and short, according to
their duration $T_{90}\lessgtr2$ sec, based on the general belief
that this observational classification is associated with a physical
one and that the two populations have different origins: long GRBs
are Collapsars and short ones are non-Collapsars (possibly arising
from neutron star mergers, but at present, this association is still
uncertain). This classification scheme is known to be imperfect due
to the large overlap in the duration distribution between the two
populations. It is also used for all detectors, although it is known
that any classification scheme depends on the detector (e.g.,
\citealt{Nakar07}). The problem with this method is that, first it
is impossible to know how trustable are results that are based on
a single classified event. Second, the level of contamination in
any studied sample is unknown. The main reason for this flawed
practice is simply the lack of a reliable and quantifiable
classification scheme. This is what we provide in this paper. Based
on a physically motivated model we have shown in an earlier study
\citep{B11a} that at short durations the Collapsar distribution is
flat, up to a typical duration of $\sim20$ sec. This enables us to
recover the non-Collapsar distribution from the overall duration
distribution and to assign probability that a burst with a given
duration and hardness is a non-Collapsar.

We carry out this analysis for three major GRB satellites, BATSE,
\Fermi and \swift. We first find the probability that a burst is a
non-Collapsar based on its duration alone, \f$(T_{90})$. We find
that it depends strongly on the observing satellite and in
particular on its spectral window. For a given duration the
probability that a BATSE burst is a non-Collapsar is larger than the
probability that a \swift burst is a non-Collapsar. A useful
threshold duration that separates Collapsars from non-Collapsars is
that where \f$(T_{90})=0.5$. We find that it is
$T_{90}=3.1\pm0.5$ sec in BATSE, $T_{90}=1.7^{+0.4}_{-0.6}$ sec in
\Fermi GBM and  $T_{90}=0.8\pm0.3$ sec in {\it Swift}.

As short GRBs are harder on average than long ones
\citep{Kouveliotou93}, it is natural to expect that GRBs with a hard
spectrum have a higher probability to be non-Collapsars than softer
ones. Thus, a better classification can be achieved by considering
the hardness, in addition to the duration. We separate the
sample of each satellite to three sub-samples based on the bursts
hardness and repeat the analysis. Not surprisingly there are fewer
non-Collapsars in the soft subgroups and more in the harder ones.
Interestingly the duration distributions of both Collapsars and
non-Collapsars depend only weakly on the hardness and only the
relative normalization between the two groups varies as we consider
subgroups of different hardness. As there are more non-Collapsars in
the hard subgroups, non-Collapsars dominate in these subgroups even
at relatively long durations. For example In the hard BATSE subgroup
the probability, \f, that a burst is a non-Collapsar remains $>0.5$
up to durations $T_{90}\simeq8$ sec. In the hard \Fermi GBM and
\swift subgroups \f$>0.5$  up to $\simeq 5$ and $\simeq3$ sec
respectively. A soft GRB, on the other hand, is more likely to be a
Collapsar. In this case \f$>0.5$ up to $T_{90}\simeq 1$ sec for
BATSE's soft subgroup, up to $T_{90}\simeq0.6$ sec in Fermi GBM and only
up to $T_{90}\simeq0.3$ in \swift's subgroups.
These values should replace the average values as dividing
durations between Collapsars and non-Collapsars, whenever hardness information
is available.
In particular for \swift,
$2.8^{+1.5}_{-1.0}$, $0.6^{+0.2}_{-0.3}$, and $0.3^{+0.4}_{-0.2}$ sec should
replace the value of $0.8\pm0.3$ sec for the hard, intermediate and soft
subgroups respectively. Our results well
agree with the overall behavior seen when comparing different
satellites. \swift's window is much softer than BATSE's and Fermi's
and the transition in \swift's overall sample between non-Collapsars
and Collapsars occurs at shorter durations relative to the other
satellites. This is a general pattern seen in both the overall
sample and in the hardness subgroups.

We find that  the transition between Collapsars and non-Collapsars
is not sharp and that there is a large overlapping region where both
Collapsars and non-Collapsars co-exist. There are short durations
Collapsars with durations  shorter than 1 sec as well as
non-Collapsars at observed durations as long as 10 sec. The
traditional method to divide bursts to ``long" and ``short"
according to a sharp observed duration criteria: $T_{90}\lessgtr2$,
introduces both ``false positive" and ``false negatives" when we
interpret duration as a proxy for a different physical origin. The
choice of the division criteria should depend on the detector's
observing windows but  it should also depend on our tolerance for
contamination  by  ``falsely" classified bursts. When interested in
Collapsars, the solution is trivial. Choosing a  conservative large
duration will eliminate a few short duration Collapsars but will results in a
sample containing practically only Collapsars. The small
number of short duration bursts makes it difficult to adopt a similar
conservative policy for them and the classification criterion should
be chosen carefully in each study. Finally, our results show clearly
that no high significance result concerning non-Collapsars can be
derived based on a single burst, which is classified according to its
high energy properties alone.

Next we examine the implication of the currently used criterion
$T_{90}\lessgtr2$ on the different satellite samples. It is
conservative for BATSE, where Only 10\% of bursts that are shorter
than 2 sec are Collapsars. One can consider a BATSE ($T_{90} \le 2$
sec) sample as reasonably free of false positives. The corresponding
fraction of ``false positives" for Fermi is higher (20\%) but still
acceptable for many purposes. However this criteria is not good for
\swift. About 40\% of \swift bursts  with $T_{90} < 2$ sec, that
have been traditionally classified and studied as SGRBs are
Collapsars. Thus the standard and commonly use sample of \swift GRBs
with $T_{90}\lessgtr2$  sec which is the source of the only sample
of well localized short GRBs is heavily contaminated with
Collapsars! This must have influenced the results of
non-Collapsars studies that are based on \swift GRBs. Interestingly,
this Collapsar contamination didn't affect qualitatively the main
conclusion concerning non-Collapsar hosts \citep{Berger09,Berger11},
namely the observation that these hosts have a wide distributions of
SFR, luminosities and metallicities and that the conclusion that the
positions of non-Collapsars has a wide spread within the host
galaxy. Such distributions are significantly different than those of
Collapsar's host. However,  quantitative features of these
distributions must have been distorted.

While the complete implications of our results on studies of
non-Collapsars is beyond the scope of this work, there are three
important points that stand out. (i) There is no convincing evidence
for high redshift non-Collapsars. All the bursts with secure
redshift that are non-Collapsars at high probability are at $z<1$.
(ii) There is no convincing evidence for beaming in non-Collapsars.
GRB 051221A is the only SGRB that show a multi-wavelength afterglow
break, that is interpreted as a jet break and is considered as the
strongest evidence for beaming in non-Collapsars. However, our
results show that the probability that this burst is indeed a
non-Collapsar is only $0.18^{+0.08}_{-0.11}$. Apparently,
non-Collapsars may or may not be beamed as far as we currently know.
(iii) The duration of a non-negligible fraction of the
non-Collapsars is $10$ s and even longer. This implies (under most
GRB models) that the central engine of these events works
continuously in the mode that produces the initial hard GRB emission
for that long. This fact should be accommodated by any model of
non-Collapsar central engine.

This research is supported by an ERC advanced research grant, by
the Israeli Center for Excellence for High Energy AstroPhysics (T.P.),
by ERC and IRG grants, and a Packard, Guggenheim and Radcliffe
fellowships (R.S.),
and by ERC starting grant and ISF grant no. 174/08 (E.N.)

\appendix
\section{Swift SGRBs}\label{App:SwiftSGRBTable}


\begin{table}\label{table.Swift_SGRBs}
\setlength{\tabcolsep}{10pt}
\begin{tabular} {|c|c|c|c|c|c|}
\hline
GRB       & $T_{90} [s] $ & PL        &  \f$^a$      &  z   &Ref\\
\hline
050202            &  0.270         &  $1.44\pm0.32$  & $0.71^{+0.27}_{-0.23}$       & &\\
{\bf 050509B}     &  0.073         &  $1.57\pm0.38$  & $0.87^{+0.04}_{-0.16}$       & 0.225 & 1\\
{\bf 050709$^b$}  &  0.07          &                 & ${0.92^{+0.02}_{-0.03}}$     & 0.161 & 2\\
{\bf050724}       &  3(96)$^\dag$  &                 &                              &  0.257 & 3\\
{\bf 050813}      &  0.450         &  $1.28\pm0.37$  & $0.57^{+0.39}_{-0.24}$       & 0.722$^*$ & 4\\
050906            &  0.258         &  $2.46\pm0.43$  &  $0.49^{+0.25}_{-0.20}$      & & \\
050925            &  0.070         &   ....$^d$....  & $0.92^{+0.02}_{-0.03}$       & & \\
051105A           &  0.093         &  $1.22\pm0.30$  & $0.85^{+0.14}_{-0.17}$      & & \\
051210            &  1.300         &  $1.06\pm0.28$  & $0.82^{+0.10}_{-0.61}$      & &\\
{\bf 051221A}     &  1.400         &  $1.39\pm0.06$  &  $0.18^{+0.08}_{-0.11}$     & 0.546 & 5\\
{\bf060121$^b$}   & 1.97           &                 & ${0.17^{+0.14}_{-0.15}}$    & $1.7\leqslant z\leqslant4.5$& 6\\
060313            &  0.740         &  $0.70\pm0.07$  & $0.92^{+0.05}_{-0.08}$      & & \\
060502B           &  0.131         &  $0.98\pm0.19$  &  $0.99^{+0.01}_{-0.16}$     & 0.287 & 7\\
{\bf 060505}      &  4.000         &  $1.29\pm0.28$  &  $0.03^{+0.29}_{-0.02}$     & 0.089 & 8\\
060614            &  6(108)$^\dag$ &                 &                             & 0.125 & 9\\
{\bf 060801}      &  0.490         &  $0.47\pm0.24$  &  $0.95^{+0.03}_{-0.05}$     & 1.131$^*$& 10\\
061006            &  0.5(123)$^\dag$&                &                             & 0.438 & 11\\
061201            &  0.760         &  $0.81\pm0.15$  &  $0.92^{+0.05}_{-0.08}$     & 0.11 / 0.087$^*$& 12 \\
061210            &  0.192(85)$^\dag$&               &                             & 0.410$^*$ & 13\\
{\bf 061217}      &  0.210         &  $0.86\pm0.30$  &  $0.98^{+0.01}_{-0.23}$     & 0.827 & 14\\
070209            &  0.090         &  $1.00\pm0.38$  &  $0.99^{+0.01}_{-0.13}$     & & \\
070406            &  1.200         &  $1.38\pm0.60$  &  $0.23^{+0.61}_{-0.13}$     & & \\
070429B           &  0.470         &  $1.72\pm0.23$  &  $0.32^{+0.26}_{-0.15}$     & 0.904 & 15\\
{\bf 070707$^c$}  &  1.1           &                 &  ${0.84^{+0.02}_{-0.03}}$   & & \\
070714A           &  2.000         &  $2.60\pm0.20$  &  $0.04^{+0.07}_{-0.02}$     & & \\
{\bf 070714B}     &  3(64)$^\dag$  &                 &                             & 0.92 & 16\\
{\bf 070724A}     &  0.400         &  $1.81\pm0.33$  &  $0.37^{+0.26}_{-0.17}$     & 0.457 & 17\\
070729            &  0.900         &  $0.96\pm0.27$  &  $0.89^{+0.06}_{-0.57}$     & & \\
070809            &  1.300         &  $1.69\pm0.22$  &  $0.09^{+0.13}_{-0.05}$     & & \\
070810B           &  0.080         &  $1.44\pm0.37$  &  $0.86^{+0.13}_{-0.16}$     & & \\
070923            &  0.050         &  $1.02\pm0.29$  &  $0.99^{+0.00}_{-0.11}$     & & \\
071112B           &  0.300         &  $0.69\pm0.34$  &  $0.97^{+0.01}_{-0.03}$     & & \\
071227            &  1.800         &  $0.99\pm0.22$  &  $0.71^{+0.15}_{-0.59}$     & 0.384& 18 \\
080121            &  0.700         &  $2.60\pm0.80$  &  $0.21^{+0.23}_{-0.11}$     & &\\
080123            &  0.8(115)$^\dag$&                &                             & &\\
080426            &  1.700         &  $1.98\pm0.13$  &  $0.06^{+0.09}_{-0.03}$     & &\\
080702A           &  0.500         &  $1.34\pm0.42$  &  $0.53^{+0.42}_{-0.23}$     & &\\
{\bf 080905A}     &  1.000         &  $0.85\pm0.24$  &  $0.88^{+0.07}_{-0.11}$     & 0.122& 19\\
\hline
\end{tabular}

\end{table}

\begin{table}
\setlength{\tabcolsep}{15.5pt}
\begin{tabular} {|c|c|c|c|c|c|c}
\cline{1-6}
GRB & $T_{90} [s] $ & PL &  \f$^a$ &  z   & Ref\\
\cline{1-6}
080919         &  0.600     &  $1.10\pm0.26$  &  $0.94^{+0.03}_{-0.47}$       & &\\
081024A        &  1.800     &  $1.23\pm0.21$  &  $0.12^{+0.59}_{-0.08}$       & &\\
081101         &  0.200     &   ....$^d$....  &  $0.85^{+0.03}_{-0.05}$       & &\\
081226A        &  0.400     &  $1.36\pm0.29$  &  $0.60^{+0.36}_{-0.24}$       & &\\
090305A        &  0.400     &  $0.86\pm0.33$  &  $0.96^{+0.02}_{-0.36}$       & &\\
090417A        &  0.072     &  ....$^d$....   &  $0.92^{+0.02}_{-0.03}$       & &\\
{\bf 090426}   &  1.200     &  $1.93\pm0.22$  &  $0.10^{+0.15}_{-0.06}$       & 2.609& 20\\
{\bf 090510}   &  0.300     &  $0.98\pm0.20$  &  $0.97^{+0.01}_{-0.29}$       & 0.903& 21\\
090515         &  0.036     &    ....$^d$.... &  $0.94^{+0.03}_{-0.07}$       & &\\
090621B        &  0.140     &  $0.82\pm0.23$  &  $0.99^{+0.01}_{-0.01}$       & &\\
090815C        &  0.600     &  $0.90\pm0.47$  &  $0.94^{+0.03}_{-0.47}$       & &\\
091109B        &  0.300     &  $0.71\pm0.13$  &  $0.97^{+0.01}_{-0.03}$       & &\\
{\bf 100117A}  &  0.300     &  $0.88\pm0.22$  &  $0.97^{+0.01}_{-0.03}$       & 0.92& 22\\
100206A        &  0.120     &  $0.63\pm0.17$  &  $0.99^{+0.01}_{-0.01}$       & &\\
100625A        &  0.330     &  $0.90\pm0.10$  &  $0.97^{+0.02}_{-0.03}$       & &\\
100628A        &  0.036     &   ....$^d$....  &  $0.94^{+0.03}_{-0.07}$       & &\\
100702A        &  0.160     &  $1.54\pm0.15$  &  $0.80^{+0.06}_{-0.20}$       & &\\
\bf{100724A}   &  1.400     &  $1.92\pm0.21$  &  $0.08^{+0.12}_{-0.04}$       & 1.288 & 23\\
101129A        &  0.350     &  $0.80\pm0.50$  &  $0.97^{+0.02}_{-0.33}$       & &\\
101219A        &  0.600     &  $0.63\pm0.09$  &  $0.94^{+0.03}_{-0.06}$       & &\\
101224A        &  0.200     &   ....$^d$....  &  $0.85^{+0.03}_{-0.05}$       & &\\
110112A        &  0.500     &  $2.14\pm0.46$  &  $0.30^{+0.26}_{-0.15}$       & &\\
110420B        &  0.084     &   ....$^d$....  &  ${0.91^{+0.02}_{-0.03}}$     & &\\
111020A        &  0.400     &  $1.37\pm0.26$  &  $0.60^{+0.36}_{-0.24}$       & &\\
111117A        &  0.470     &  $0.65\pm0.22$  &  $0.96^{+0.03}_{-0.05}$       & &\\
111126A        &  0.800     &  $1.10\pm0.30$  &  $0.91^{+0.05}_{-0.54}$       & &\\
\cline{1-6}
\end{tabular}\\
\begin{tabular} {@{}lcccccc}
 \multicolumn{7}{l}{\footnotesize $^a$ \swift GRBs with a single power-law spectral fit are assigned a probability \f$(T_{90},PL)$}\\
 \multicolumn{7}{l}{\footnotesize \hspace{6pt}  Other GRBs can only be assigned a probability \f$(T_{90})$.} \\
 \multicolumn{7}{l}{\footnotesize $^b$ A GRB detected by HETE, \f$(T_{90})$ is estimated using the \swift probability function}\\
 \multicolumn{7}{l}{\footnotesize $^c$ A GRB detected by Integral, \f$(T_{90})$ is estimates using the BATSE probability function}\\
 \multicolumn{7}{l}{\footnotesize $^d$ The spectral fit of the $\gamma$-ray photons is a power-law with an exponential cutoff,}\\
 \multicolumn{7}{l}{\footnotesize \hspace{9pt} \f$(T_{90},PL)$ cannot be calculated for this burst and \f$(T_{90})$ is used instead.} \\
 \multicolumn{7}{l}{\footnotesize $^\dag$ A GRB with an extended soft emission, no \f is assigned.}\\
 \multicolumn{7}{l}{\footnotesize $^*$ Unsecure redshift, based on an association of a galaxy within the XRT error circle.}\\
 \multicolumn{7}{l}{\scriptsize Redshift
  references: (1)\citet{Prochaska05a,Gehrels05}; (2)\citet{Villasenor05,Fox05}; }\\
 \multicolumn{7}{l}{\scriptsize (3)\citet{Berger05,Prochaska05b}; (4)\citet{Gehrels05,Berger05b,Foley05}; }\\
 \multicolumn{7}{l}{\scriptsize (5)\citet{BergerSod05,Soderberg06b}; (6)\citet{Ugarte06,Levan06};}\\
 \multicolumn{7}{l}{\scriptsize (7)\citet{Bloom06}; (8)\citet{Ofek06,Levesque07}; (9)\citet{Price06,Fugazza06};}\\
 \multicolumn{7}{l}{\scriptsize (10)\citet{Cucchiara06}; (11)\citet{Berger07}; (12)\citet{Berger06,Berger07b}; (13)\citet{Cenko06};}\\
 \multicolumn{7}{l}{\scriptsize (14)\citet{Berger06b}; (15)\citet{Perley07}; 16)\citet{Graham07}; (17)\citet{Cucchiara07,Covino07}; }\\
 \multicolumn{7}{l}{\scriptsize (18)\citet{D'Avanzo07,Berger07c}; (19)\citet{Rowlinson10}; (20)\citet{Levesque09};}\\
 \multicolumn{7}{l}{\scriptsize \citet{Thoene09}; (21)\citet{Rau09};  (22)\citet{Fong11}; (23)\citet{Thoene10}}\\
  \end{tabular}

\end{table}

\end{document}